\documentclass[a4paper,11pt]{article}
\pdfoutput=1 

\usepackage{jcappub} 

\usepackage[T1]{fontenc} 
\usepackage{subcaption} 
\usepackage{graphicx}
\usepackage{multirow}
\usepackage{epstopdf}
\usepackage[normalem]{ulem}

\title{\boldmath Principal component analysis of the primordial tensor power spectrum}

\author[]{Paolo Campeti,}
\author[]{Davide Poletti,}
\author[]{and Carlo Baccigalupi}

\affiliation[]{SISSA - Scuola Internazionale Superiore di Studi Avanzati,\\Via Bonomea 265, 34136, Trieste, Italy}
\affiliation[]{IFPU - Institute for Fundamental Physics of the Universe, \\Via Beirut 2, 34014, Trieste, Italy}
\affiliation[]{INFN - National Institute for Nuclear Physics, Sezione di Trieste,\\ Via Valerio 2, 34127, Trieste, Italy}

\emailAdd{paolo.campeti@sissa.it}
\emailAdd{davide.poletti@sissa.it}
\emailAdd{carlo.baccigalupi@sissa.it}

\abstract{We study how the shape of the spectrum of primordial gravitational waves can be constrained by future experiments looking at the B-mode of the Cosmic Microwave Background (CMB) polarization. We implement a Principal Component Analysis (PCA) including the effects of diffuse foreground residuals, following component separation, in the uncertainty of CMB angular power spectra, and taking into account the gravitational lensing by Large Scale Structure. We perform our study by considering the capabilities of future B-mode CMB experiments such as LiteBIRD, the Simons Observatory (SO) and Stage-IV (CMB-S4), in particular in detecting deviations of the primordial tensor spectrum from the scale-invariant behavior.
We find that diffuse foreground residuals impact substantially both the derivation of the PCA basis and the corresponding constraining power, in all cases. In particular, depending on which experimental specifications and which value $r$ of tensor-to-scalar ratio for cosmological perturbations are considered, adding foregrounds residuals can determine an increase as large as a factor $\sim 4$ both on the uncertainty on $r$ and on the recovery of the PCA modes. 
We study the limitations of the methodology, including the effect of physicality priors on the PCA, which we quantify via a Monte Carlo Markov chain (MCMC) analysis of the combined cosmological and PCA power spectrum parameter space.
}

\arxivnumber{1905.08200}
\begin{document}
\maketitle
\flushbottom

\section{Introduction}
\label{sec:intro}
Tensor perturbations in cosmology can be produced in the very early Universe along with scalar and vector ones, and constitute the Cosmic Gravitational Background (CGB) of radiation. According to the inflationary paradigm, semi-classical graviton production occurs as Fourier components of the fluctuation tensor field in the Friedmann-Lema\^{\i}tre Robertson-Walker (FLRW) metric cross the horizon, recording the expansion rate at that epoch. Cosmological perturbations carry detailed information concerning the physics of the early Universe, and, in the standard single-field, slow-roll inflationary scenario, of the effective inflaton field and its potential and dynamics \citep{liddle_and_lyth}: perturbations in the CGB are Gaussian, and their power spectrum is characterized through a relative amplitude with respect to scalar perturbations at cosmological scales, $r$, and a spectral index, $n_{T}$, specifying its power law as a function of the wavenumber $k$. However, while robust measurements of scalar perturbations have been achieved through modern cosmological observations of the Cosmic Microwave Background (CMB) anisotropies and Large Scale Structure (LSS) \citep[see e.g.][and references therein]{planck_2018}, no detection exists of primordial vector or tensor modes, so that their amplitude and statistics has to be ascribed to unknown physics. 

Cosmological gravitational waves are known to source the divergenceless component of the CMB polarization, the B-mode, with a maximum contribution on large scales, hundreds of comoving Mega-parsecs (Mpc), where they are boosted by rescattering onto electrons freed by cosmic reionization, and on the degree angular scale, corresponding to the cosmological horizon at recombination. On smaller scales, they are dominated by the contribution of divergence (E-)modes into CMB polarization leaking into B due to gravitational lensing deflection. 
Due to the insight into the physics of the very Early Universe that the discovery of the CGB would give, several experiments are operating and others are planned for the next decade. B-mode measurements from gravitational lensing have been achieved by Planck \citep{planck_b_modes}, Background Imaging of Cosmic Extragalactic Polarization 2 (BICEP2) \citep{bicep_b_modes}, POLARBEAR \citep{polarbear_b_modes}, the Atacama Cosmology Telescope \citep[(ACT), see][]{act_b_modes} and the South Pole Telescope \citep[(SPT), see][]{spt_b_modes}. The only detection of degree scale B-modes of primordial origin has been claimed by the BICEP2 experiment \citep{bicep_ref_2014}, but the uncertainty due to possible contamination from diffuse Galactic foregrounds did not allow to confirm the origin of the signal  \citep{bicep_planck_ref}. Over the incoming years, the Simons Array \citep[see][]{sa_ref}, the Simons Observatory \citep[hereafter SO, see][]{so_ref}, the Stage-IV network of ground-based observatories \citep[hereafter CMB-S4, see][]{cmbs4_book}, and the Light satellite for the study of B-mode polarization and Inflation from cosmic microwave background Radiation Detection \citep[hereafter LiteBIRD, see][]{litebird_ref} will be observing the microwave sky at multiple frequencies in order to control the diffuse foreground emissions, on regions ranging from a few percent fraction to the whole sky, and with increasing sensitivity. 

In addition to the standard prediction of single-field, slow-roll inflation models, several other production mechanisms have been studied as possible sources of primordial B-modes; among them, massive gravity inflation \citep{domenech_etal_2017}, open inflation \citep{yamauchi_etal_2011}, the SU(2)-axion model \citep{Dimastrogiovanni_etal, Adshead_etal, Maleknejad_etal}, modified gravity models with speed of gravitational waves different from light speed \citep{raveri_etal_2014}, models of inflation with topological defects/cosmic strings \citep{lizarraga_etal_2014}, rolling axion \citep{namba}, high-scale inflation \citep{baumann}, multifield inflation \citep{price} and others. In order to be able to distinguish different physical mechanisms it is necessary to be able to quantify deviations of the measured spectrum from the power-law inflationary prediction, as it is done in \citep{hiramatsu_etal_2018} for the first time for the reconstruction of the tensor power spectrum. 
This work aims at providing a Principal Component Analysis (PCA) of the primordial tensor spectrum, similar to what has been done for scalar modes prior to the Planck mission \citep[see ][and references therein]{leach_2006}. The PCA formalism allows to identify and study eigenvectors of the Fisher matrix associated to the primordial perturbation spectra, assessing the modulation in sensitivity of an experiment on the different cosmological perturbation scales, and determining where features in the tensor power spectrum can be probed more efficiently. This technique has been applied in many different contexts, i.e. the scalar primordial power spectrum \citep{hu_okamoto_2004, leach_2006}), the reconstruction of the inflaton potential \citep{kadota_etal_2005}, the process of reionization \citep{hu_holder_2003}, the dark energy equation of state \citep{huterer_starkman_2003}, weak lensing \citep{munshi_kilbinger_2006}, the optimal binning of the primordial power spectrum \citep{paykari_jaffe}, the search of inflationary and reionization features in the Planck data \citep{obied_etal_2018} and many others. Although with different procedure and final goals, PCA has also been applied to the tensor power spectrum in a recent work \citep{farhang_sadr_2018}. We expand these analyses and set rather different perspectives for our study, $(i)$ by including, for the first time as a major aspect and in a coherent manner, the contribution from residual diffuse foreground contamination (Section \ref{sec:foregrounds}), using the most recent techniques of foreground separation and cleaning via Maximum Likelihood Parametric Fitting implemented into the ForeGroundBuster (FGBuster) publicly available code\footnote{See github.com/fgbuster/fgbuster and reference therein.}; $(ii)$ by deriving robust expectations concerning the sensitivity of three forthcoming CMB probes (Sections \ref{sec:litebird} and \ref{sec:SOandS4}) and modelling these experiments as realistically as possible (including $1/f$ noise when available); $(iii)$ by investigating through a Monte Carlo Markov chain (MCMC) analysis the limitations of this technique, namely the impact of the physicality priors (which prevent the exploration of forbidden regions where the power spectrum assumes negative, and thus unphysical, values) on the derived constraints, the residual correlations between parameters and the departures from Gaussian behaviour (Sections \ref{sec:mcmc} and \ref{sec:limitations}). 

The paper is organized as follows. In Section \ref{sec:standardmodel} we give the necessary definitions, and in Section \ref{sec:PCA} we review the general PCA algebra. In Section \ref{sec:foregrounds} we study and include the diffuse astrophysical foregrounds in our analysis, assessing the level of contamination and its role in the PCA, in Section \ref{sec:foregrounds:contamination} and \ref{sec:foregrounds:impact}, respectively. In Section \ref{sec:results} we predict and study the capability of future satellite (\ref{sec:litebird}) and ground-based (\ref{sec:SOandS4}) probes of constraining deviations from the inflationary CGB spectrum; in Section \ref{sec:earlyuniversemodels} we give explicit examples of deviations from the inflationary predictions and how PCA can measure and characterize those; Section \ref{sec:limitations} is dedicated to the study of the limitations and caveats imposed by the PCA method in the context of the tensor power spectrum. Finally, in Section \ref{sec:conclusions} we outline our conclusions and future perspectives. 

\section{Power spectra, parameters and observations}\label{sec:standardmodel} 
In this section, we briefly review the standard formalism for scalar and tensor power spectra of primordial perturbations, as well as for the angular power spectra in the CMB, and we define our fiducial cosmological model. Moreover, we define the nominal specifications of the CMB B-mode probes we will consider throughout the analysis. 

According to single field, slow-roll inflationary scenario, quantum vacuum fluctuations excite cosmological scalar, vector and tensor perturbations. While vector modes decay, scalar and tensor modes in the metric stay constant and we focus on them in the following. The scalar curvature spectrum can be parametrized in the usual way as a power-law spectrum,
${\mathcal P}_{{\mathcal 
		R}}(k)= A_{s}
\left(k / k_0\right)^{n_{{s}}-1}\,$,
where $A_{s}$ is the amplitude of scalar perturbations, $n_s$ the scalar spectral index and $k$ the wavenumber of the perturbation. $k_{0}$ is a pivot-scale, which we set to $0.002$ $Mpc^{-1}$. No information is stored in the higher order statistics, as the vacuum fluctuations are assumed to be Gaussian distributed. The presence of matter fields during inflation changes this picture \citep[see][]{Agrawal_etal}. 
Similarly, the tensor power spectrum can be written 
4
 using the same standard power-law parametrization, 
${\mathcal P}_{T}(k)= A_{T}
	\left(k/k_0\right)^{n_T}\,$,
 where $A_{T}$ is the amplitude of tensor perturbations and $n_T$ the tensor spectral index. We define also the tensor-to-scalar (hereafter $T/S$) ratio $r$ as
$r=A_{T} / A_{s}$.
Currently there are only upper limits available on r, in particular $r<0.06$  
at $95\%$ CL from the combination of B-mode polarization data from BICEP2- Keck \citep{bicep2_2018} and Planck 2018 data \citep{planck_2018}. 

The actual observables are not scalar or tensor power spectra, but
the angular power spectra $C_{\ell}^{XX'}$ of CMB anisotropies, where $X$ are generic labels for the total intensity of linear polarization. They are defined using the two-point correlation function of the spherical harmonic coefficients 
$\langle a_{\ell m}^{X*} a_{\ell' m'}^{X'} \rangle = \delta_{\ell\ell'}\delta_{mm'} 
C_\ell^{XX'}\,$,
where $X,X' \in \{T,E, B\}$, representing respectively the total intensity (T), gradient (E) and curl (B) modes of the CMB polarization \citep{kamionkowski_kosowski_stebbins_1997,seljak_zaldarriaga_1997} and $a_{\ell m}^{X}$ are the multipole moments of total intensity and polarization fluctuations. 
We provide now the link between the observable angular power spectra and the primordial one. First of all, we remind that each angular power spectra has uncorrelated scalar and tensor contributions, that is
$C_{\ell}^{XX',\,prim}= C_{\ell,s}^{XX'}+ C_{\ell,t}^{XX'}$.
The	scalar and tensor contribution can be written in terms of the primordial power spectra as
\begin{equation}
C_{\ell, x}^{XX'} = \frac{2 \pi}{\ell(\ell+1)} \int d \ln k\,\mathcal{P}_{y}\left(k\right) 
T^{X}_{\ell, y}\left(k\right) T^{X'}_{\ell, y}\left(k\right)\,,
\end{equation}
where for the scalar case $X,X'=\{T,E\}$, $x=\{s\}$ and $y=\{\mathcal{R}\}$, while for the tensor one $X,X'=\{T,E, B\}$, $x=\{t\}$ and $y=\{T\}$. The $T^{X}_{\ell, y}$ are the scalar or tensor transfer functions; they depend on cosmological parameters as we review next, 
and are obtained from the solution of the Boltzmann equations. We compute them using the publicly available code CAMB (Code for Anisotropies in the Microwave Background \citep{camb}) .

\subsection{Fiducial cosmology}\label{sec:fiducial_cosmology}
Our fiducial cosmological model is a flat $\Lambda CDM$ with the six parameters fixed by
the most recent observations \citep{planck_2018}. We choose the best fit cosmological parameters from Table 2 in the quoted reference, obtained from the $TT$, $TE$, $EE$ spectra and also including the large scale polarization (labeled as {\it low\ E}) and gravitational lensing. The model is made by 4 parameters expressing background quantities in a flat FLRW Universe, specifically the abundance of particles in the standard model ($\Omega_{b} h^{2}=0.02237$), CDM ($\Omega_{c}h^{2}=0.12$), reionization optical depth ($\tau=0.0544$), amplitude of the Hubble constant today ($H_{0}=67.36$); 2 parameters define the power spectrum of scalar perturbations, namely its amplitude ($A_{s}=2.1\times 10^{-9}$), and scalar spectral index 
($n_{s} =0.9649$). For what concerns the tensor power spectrum, we consider three cases, corresponding to $r = 0$, $r=0.001$ and $r=0.01$.
The value $r=0.001$ is particularly relevant from the point of view of both observation and theory: it is close to the limit sensitivity of future B-mode probes and the prediction of the original Starobinsky model of inflation
\citep[see][and references therein]{planck_2018_inflation}. The $r=0.01$ case would be a strong signal within reach by the operating probes in the near future. For both models with $r>0$, we assume a scale-invariant spectrum with tensor spectral index $n_T = 0$. 

\subsection{Instrumental specifications}\label{sec:instrumental_specifications}
As we anticipated, we will consider several future CMB B-mode probes, representing the ongoing projects with ultimate sensitivity for $r$ and $n_{T}$. The relevant parameters for us are represented by the instrumental sensitivity, usually given in $\mu K$-arcmin, the full width at half maximum (FWHM), the sky fraction considered, $f_{sky}$, as well as the relevant multipole range in the angular domain. 
As anticipated in the introduction. we will consider the LiteBIRD satellite\footnote{See litebird.jp/eng.} designed in order to have as primary goal the CGB observations from space \citep{litebird_ref}. For what concerns ground probes, we consider the SO {\citep{so_ref}; in particular, we consider the small aperture telescopes (SATs), since they will gather most of the information on the primordial B-mode power spectrum. Finally, we also study the case of the ultimate network of ground-based probes, the CMB-S4 
\citep{cmbs4_book}.\footnote{For CMB-S4 we also use specifications from the websites \url{https://www.nsf.gov/mps/ast/aaac/cmb_s4/report/CMBS4_final_report_NL.pdf} and \url{https://cmb-s4.org/wiki/index.php/Survey_Performance_Expectations}.\label{websites}}  
We report in Table \ref{table:instrumentalspec} the instrumental specifications for these three experiments. The sensitivity reported is for polarization measurements. The $\ell$ range for LiteBIRD is given by $\ell_{min} = 2$ and $\ell_{max} = 1350$, for SO and CMB-S4 is instead $\ell_{min} = 30$ and $\ell_{max} = 4000$.
The Table also resports the sky fraction $f_{sky}$ and the delensing factor $\lambda$ (defined in Section \ref{sec:lensing_noise_foregrounds}) employed for each experiment. For SO and CMB-S4 we additionally take into account $1/f$ noise, adopting the optimistic case of \citep{so_ref} for SO and using the specifications contained in the websites in footnote \ref{websites} for CMB-S4. For LiteBIRD, the reported specifications were taken during the Phase A1 process (Hazumi 2019, private communication) and may slightly differ from the definitive ones. 

Following \citep{stompor_etal_2016} and \citep{so_ref}, the instrumental noise model for a given experiment and a given frequency $\nu$ is 
\begin{equation}\label{eq:noise}
N_{\ell, \nu}^{XX}=\left[w_{X,\nu}^{-1}\exp\left(\ell(\ell+1)\frac{\theta^{2}_{FWHM,\nu}}{8\log 2}\right)\right]\cdot \left[1+\left(\frac{\ell}{\ell_{knee}}\right)^{\alpha_{knee}}\right],
\end{equation}
where $w_{X,\nu}^{-1/2}$ is the sensitivity of the experiment (white noise level) in the frequency channel $\nu$ in $\mu K$-rad, $\theta_{FWHM,\nu}$ represents the beam size in radians and $\alpha_{knee}$ and $\ell_{knee}$ parametrize the $1/f$ noise contribution for each frequency channel. Moreover, we assume perfect polarization efficiency, so that $w_{E}^{-1/2} = w_{B}^{-1/2} = \sqrt{2}w_{T}^{-1/2}$.

\begin{table}
	\centering
	\footnotesize{
	\begin{tabular}{|c|c|c|c|c|c|}
		\hline \hline
		Experiment
		& Frequency
		& Sensitivity 
		& FWHM 
		& $\ell_{knee}$
		& $\alpha_{knee}$
		\\
		
		& [GHz]
		& [$\mu$K-arcmin]
		&  [arcmin]
		& 
		& 
		\\
		\hline	\hline
		
		&	40.0 & 36.1 & 69.2 & & \\
		&	50.0 & 19.6 & 56.9 & & \\
		&	60.0 & 20.2 & 49.0 & & \\
		&	68.0 & 11.3 & 40.8 & & \\
		&	78.0 & 10.3 & 36.1 & & \\
		&	89.0 & 8.4  & 32.3 & & \\
	\textbf{LiteBIRD}	&	100.0 & 7.0 & 27.7 & n/a & n/a\\
	($f_{sky}=0.6$;	&	119.0 & 5.8 & 23.7 & & \\
	$\lambda=0.8$)	&	140.0 &   4.7    &  20.7 & & \\
		&	166.0 &   7.0    &    24.2   & &  \\
		&	195.0 &    5.8   &     21.7     & & \\
		&	235.0 &   8.0    &   19.6  & & \\
		&	280.0 &    9.1   &     13.2    & &  \\
		&	337.0 &    11.4   &    11.2       & & \\
		&	402.0 &     19.6  &    9.7  & &     \\
			\hline\hline
			
		& 27.0 & 35.3 & 91.0 &  15 & -2.4 \\
	\textbf{SO (SATs)}	& 39.0 & 24.0 & 63.0 &  15 & -2.4\\
	($f_{sky}=0.1$;	 & 93.0 & 2.7  &  30.0   & 25 & -2.6\\
$\lambda=0.5$)	& 145.0 & 3.0 &  17.0  & 25 & -3.0\\
 	& 225.0 & 5.9 &  11.0  & 35 & -3.0\\
		& 280.0 & 14.1 &   9.0  & 40 & -3.0\\
		\hline\hline
		
		& 20.0 & 14.0 &  11.0 &   200 & -2.0\\
		& 30.0 & 8.7 &   76.6 &  50 &-2.0\\
		& 40.0 & 8.2 &    57.5        & 50&-2.0  \\
	\textbf{CMBS-S4}	& 85.0 & 1.6 &   27.0 &   50 &-2.0\\
($f_{sky}=0.03$;		& 95.0 & 1.3 &  24.2 &   50 &-2.0\\
$\lambda=0.1$)		& 145.0 & 2.0 &   15.9 &  65 &-3.0\\
		& 155.0 & 2.0 &   14.8 &   65 &-3.0\\
		& 220.0 & 5.2 &   10.7 &   65 &-3.0\\
		& 270.0 & 7.1 &   8.5 &   65 & -3.0\\
		\hline\hline
	\end{tabular}}
	\caption{Instrumental specifications for the LiteBIRD, SO and CMB-S4 experiments.}
	\label{table:instrumentalspec}
\end{table}

\subsection{Astrophysical and instrumental contributions to the observed power spectrum}\label{sec:lensing_noise_foregrounds}
In the following, we will use $C_{\ell}^{XX'}$ for the total observed power spectrum
\begin{equation}\label{eq:total_signal}
C_{\ell}^{XX'} = C_{\ell}^{XX',\,prim}+C_{\ell}^{XX',\,noise} + \lambda C_{\ell}^{XX',\,lens} + C_{\ell}^{XX,\,fgs}\,,
\end{equation}
where $C_{\ell}^{XX',\,prim}$ is the primordial angular power spectrum, 
$C_{\ell}^{XX',\,noise}$ is the contribution from instrumental noise (including the $1/f$ noise term and the noise degradation after component separation as we will define in Section \ref{sec:foregrounds:contamination}), $\lambda C_{\ell}^{XX',\,lens}$ represents the lensing term and $C_{\ell}^{XX,\,fgs}$ is the residual contamination by polarized diffuse foregrounds, which will be described in detail in Section \ref{sec:foregrounds}. 

The contribution from gravitational lensing acts as a contaminant when estimating the primordial CMB power spectrum. This term can be modelled, thus removing the bias in the observed power spectrum. However, the associated cosmic variance will still limit the constraints, especially on B-modes. By performing the so-called delensing in the map-domain \citep{delensing1,delensing2, delensing3, delensing4} the lensing correction to the observed power spectrum and the associated cosmic variance can be suppressed. We model the result of this operation directly in the power spectrum domain, suppressing $C_{\ell}^{XX',\,lens}$ by a constant factor $\lambda$. $\lambda=0$ means that the lensing contribution is completely removed and $\lambda=1.0$ means that no delensing is performed. The effectiveness of delensing depends on the noise level and the resolution of of the experiment, we take $\lambda = 0.8$ for LiteBIRD \citep[see][]{namikawa_etal_2015}, $\lambda = 0.5$ for SO \citep{so_ref} and $\lambda = 0.1$ for CMB-S4 \citep[see][]{cmbs4_book}. As for $C_{\ell}^{XX',\,prim}$,  $C_{\ell}^{XX',\,lens}$ is evaluated with CAMB.

\section{Principal component analysis of tensor power spectrum}\label{sec:PCA}
In this section we briefly review the formalism of Fisher information matrix (Section \ref{sec:fishertheory}) and the PCA (Section \ref{sec:PCAtheory}). We apply it to the tensor power spectrum and highlight the new aspects related to this context (Section \ref{sec:mcmc}).

\subsection{Fisher information matrix for tensor power spectrum}\label{sec:fishertheory}
In order to discretize the tensor power spectrum $\mathcal{P}_{T}$, following \citep{hu_okamoto_2004}, we write \begin{equation}\label{eq:expansion}
\mathcal{P}_{T}(k) = A_{s} \sum_i p_i W_i(\ln k)\,,
\end{equation}
where 
$W_i$ is the discretization window function, which we choose to be a triangle
\begin{eqnarray}
W_i(\ln k) =  {\rm max}\left(1- \bigg|\frac{\ln k -\ln k_i}{\Delta \ln k}\bigg|,0\right) \,,
\end{eqnarray}
and
\begin{equation}
\Delta \ln k= \ln k_{i+1} -\ln k_i \,,
\end{equation}
is the discretization constant. In this discrete representation, the derivative of the $C_{\ell}^{XX'}$ with respect to the power spectrum parameters $p_i$ becomes 
\begin{equation}\label{eq:derivative}
D_{\ell i}^{XX'} = \frac{\partial C_\ell^{XX'}}{ \partial p_i}\bigg|_{\rm fid}
	=\frac{2 \pi}{ \ell(\ell+1)} A_{s} \int d\ln k\,
	T^{X}_\ell(k)\, T^{X'}_\ell(k)\,  W_i(\ln k)\,.
\end{equation}
We choose the $k$ range $10^{-4}< k < 0.2\, {\rm Mpc}^{-1}$, which comfortably contain the scales to which the CMB power spectrum is sensitive to. The discretization scale is chosen to be $\Delta \ln k=0.05$, sharper features are smeared out because of geometrical projection effects and lensing \citep{hu_okamoto_2004}. 

The Fisher information matrix for a Gaussian field on the sphere is \citep[see, e.g.,][]{tegmark_1996}
\begin{equation}
\label{eq:x}
F_{ij} = f_{sky}
\sum_{\ell=2}^{\ell_{\rm max}}
\frac{2 \ell +1}{2} 
{\rm Tr} \left[{\bf D}_{\ell i} 
{\bf C}_{\ell}^{-1}
{\bf D}_{j \ell} {\bf C}_{\ell}^{-1}\right]\,,
\end{equation}
where we are approximating the loss of information due to the partial celestial coverage with a factor proportional to the covered sky fraction $f_{sky}$, ${\bf C}_{\ell}$ is the matrix 
	\begin{equation}
	{\bf C}_{\ell}=\left( \begin{array}{ccc}
	C_{\ell}^{TT} & C_{\ell}^{TE} & 0 \\
	C_{\ell}^{TE} & C_{\ell}^{EE} & 0 \\
	0 & 0 & C_{\ell}^{BB} \end{array} \right)\,,
	\end{equation}
and ${\bf D}_{i \ell}$ is its derivative with respect to $p_{i}$.

 As emphasized in \citep{hu_okamoto_2004}, the lensing contribution $C_{l}^{XX',\, lens}$ contains the product between $C_{\ell}^{BB}$, which depends on the primordial tensor power spectrum parameters, and $C_{\ell}^{\phi\phi}$. Therefore, when performing the derivatives with respect to the power spectrum parameters $p_{i}$ in \eqref{eq:derivative}, we take into account this dependence. The lensing potential, instead, does not depend on the tensor power spectrum, so does not contribute to the $D_{\ell}^{XX'}$.

We address also the possible issue of degeneracies between the effect that the primordial power spectrum and the other cosmological parameters can have on the $C_{\ell}$s. In particular, following \citep{hu_okamoto_2004} and \citep{leach_2006}, the Fisher matrix $F_{\mu\nu}$ for both power spectrum and cosmological parameters can be written as a block matrix
\begin{equation}
 F_{\mu\nu}=\left( \begin{array}{cc}
 F_{ij} & {\bf B} \\
{\bf B}^{\bf T} &  F_{ab} \end{array} \right)\,,
\label{eq:block}
\end{equation}
where $F_{ab}$ is the block for the cosmological parameters \{$A_{s}$, $n_{S}$, $\tau$, $\Omega_{b}h^{2}$, $\Omega_{D}h^{2}$, $H_0$\}, $F_{ij}$ is the one for the power spectrum parameters and $\mathbf{B}$ contains the cross terms between power spectrum and cosmological parameters.
We emphasize that $r$ is not included in the cosmological parameters so that the primordial tensor power spectrum is entirely defined by the $p_i$ coefficients in Eq.~\ref{eq:x}. We study the effect of the inclusion of $r$ in Section \ref{sec:limitations}.
 Inverting Eq.~\ref{eq:block} we obtain the covariance matrix $C_{\mu\nu} = F_{\mu\nu}^{-1}$, whose upper diagonal block $C_{ij}$ is the covariance matrix for the power spectrum parameters orthogonalized with respect to the cosmological parameters and the corresponding Fisher matrix would be $ F^{\,orth}_{ij} = (C_{ij})^{-1}$. Performing a block-wise inversion of the matrix in Eq.~\ref{eq:block} \citep[see, e.g.,][]{numerical_recipes} we get   
\begin{equation}\label{eq:block_inversion}
F^{\,orth}_{ij}=F_{ij}-{\bf B}\,F_{ab}^{-1}\,{\bf B}^{\bf T}\,,
\end{equation}
where the first term is the original information content on the primordial power spectrum and the second term expresses the information loss due to the degeneracy with the cosmological parameters.

\subsection{Principal component analysis}\label{sec:PCAtheory}

PCA \citep[see e.g.][]{hu_okamoto_2004,paykari_jaffe,munshi_kilbinger_2006} aims at identifying the uncorrelated variables and ranking them according to their uncertainty. In practice, since  we assume the covariance matrix to be the inverse of the Fisher information, it consists in performing the singular-value decomposition 
		\begin{equation}
		\textbf{F}=\textbf{S}^{\textbf{T}}\,\textbf{E}\,\textbf{S}\,, 
		\end{equation}
where the rows of $\textbf{S}$ are the eigenvectors of $\textbf{F}$, $\textbf{E}=diag({\bf e})$ and $e_{i}$ are the eigenvalues of $\textbf{F}$ ordered from the largest to smallest. The PCA produces in this way a new set of parameters $m_a$ -- called PCA amplitudes -- that are linear combinations of the original parameters $p_i$
\begin{equation}
\textbf{m}=\textbf{S}\,\textbf{p}\,.
\end{equation}
The covariance of these new parameters is $\textbf{E}^{-1}$ and therefore they are uncorrelated and the uncertainty on their determination is given by
\begin{equation}
\sigma_{a}= e_{a}^{-1/2}\,,
\label{eq:uncertainties}\end{equation}
where the first PCA amplitude $m_1$, corresponding to the largest eigenvalue, is the best-measured component, while the last PCA amplitude
$m_n$, corresponding to the smallest eigenvalue, is the worst-measured component. In summary, PCA finds a natural basis for the free parameters for a given experimental configuration and tells us the linear combinations of the original parameters that can be determined best.

Another goal of PCA is to compress the information. Suppose that we want to consider only a fixed number of linear combinations of the parameters, the first PCA amplitudes are the choice that retains the largest fraction of the total information. In many cases -- which include ours -- most of the information is retained in the first few  PCA modes. We can determine the number of PCA modes that is worth keeping in our analysis by plotting the  information fraction retained in first $N$ modes (e.g. Figure \ref{fig:information_LiteBIRD}), as we will describe in Section \ref{sec:results}.

\subsection{PCA modes for model testing}\label{sec:mcmc}
The $\mathcal{S}_{a}(k)$, with $a$ lower then some $N$, can be used as a basis for the primordial tensor power spectrum. Of course, they span a subspace of all possible functions -- the modes to which the given experimental configuration is sensitive to. This basis can be used to probe the detectability of specific theoretical models that predict features in the tensor power spectrum (see Section \ref{sec:intro}). We discuss here two approaches that we exploit and compare later in the paper.

We already have the Fisher uncertainty on the $m_a$ coefficients from the construction of the PCA basis. Therefore, given a theoretical power spectrum $\mathcal{P}_{model}$, it is natural to forecast how detectable it is by first projecting it over the PCA modes
\begin{equation} \label{eq:projection}
m_a =\int d \ln k \,{\mathcal S}_a(k) \mathcal{P}_{model}(k) \,,
\end{equation}
and then evaluating the probability of getting a value higher than $\sum_{a = 1}^{N} (m_a / \sigma_a)^2$ from a $\chi^2$ distribution with $N$ degrees of freedom. This significance forecast is both extremely fast and easy to perform because it neither involve any additional run of Boltzmann codes nor require any likelihood sampling. It is therefore particularly attractive for studying large sets of inflationary models and probing their parameters space.  

This approach is essentially a Fisher estimation that has notable caveats. First, the uncertainties are lower bounds that are not guaranteed to be reached. Second,  
this formalism -- including the way we constructed the PCA modes -- is insensitive to the physicality prior $P_{T} > 0$: it is based on the curvature of the likelihood with respect to the PCA amplitudes but the likelihood is not differentiable around $P_{T} = 0$.
This is formally true also for $r$ or each entry of the $\mathbf{p}$ vector, but the physicality prior translates into $r > 0$ and $p_i > 0$ and therefore their one-sided derivatives and curvatures are well defined.

Since this is not true for PCA amplitudes, we also consider another route to model testing. 
We modify the CosmoMC \citep{lewis_bridle_2002,lewis_2013} package for MCMC, redefining the tensor power spectrum \citep{leach_2006}  as  
\begin{equation}\label{eq:tensor_mcmc}
 {\mathcal P}_{T}(k)= A_{s} \sum_{a=1}^{N} m_a \mathcal{S}_a(k)\,,
 \end{equation}
where the $\mathcal{S}_a$ are the same PCA modes defined in the previous approach. We then fit the full set of parameters \{$m_{1}$,..., $m_{N}$, $A_{s}$, $n_{s}$, $\tau$, $\Omega_{b}h^{2}$, $\Omega_{D}h^{2}$, $H_{0}$\} to a given angular power spectrum assuming flat priors. 
Clearly, if the best fit is consistent with all the PCA amplitudes $m_a$ being consistent with zero, no deviation from scale-invariance is detected. This approach can account for the full complexity of the posterior distribution and, in particular, the physicality prior ${\mathcal P}_{T}(k) > 0$.

As we anticipated in Section \ref{sec:intro}, several examples of theoretical models that predict features in tensor power spectrum have been studied in the literature. It is beyond the scope of this paper to investigate specifically these scenarios since our focus is on the applicability and limitations of the PCA to express constraints on the tensor power spectrum. However, in Section \ref{sec:earlyuniversemodels}, we apply this formalism to a toy-model of red-tilted spectrum.

\section{The role of diffuse astrophysical foregrounds}\label{sec:foregrounds}
In the estimation of the Fisher matrix we include the uncertainty due to the removal of diffuse foregrounds, in addition to the ones from instrumental noise, lensing and cosmic variance. This contribution has not been considered up to now in the PCA literature and in particular for the analysis concerning the tensor spectrum. However, it is well known that diffuse foregrounds are the predominant source of uncertainty on large scale B-mode polarization \citep[see e.g.][ and reference therein]{planck_2018_diffuse_component_separation}. In this section we first explain how we model this uncertainty and then show that including foregrounds is necessary, especially when the experimental configuration provides access to the largest angular scales, including the reionization bump.


\begin{figure}
    	\includegraphics[scale=0.61]{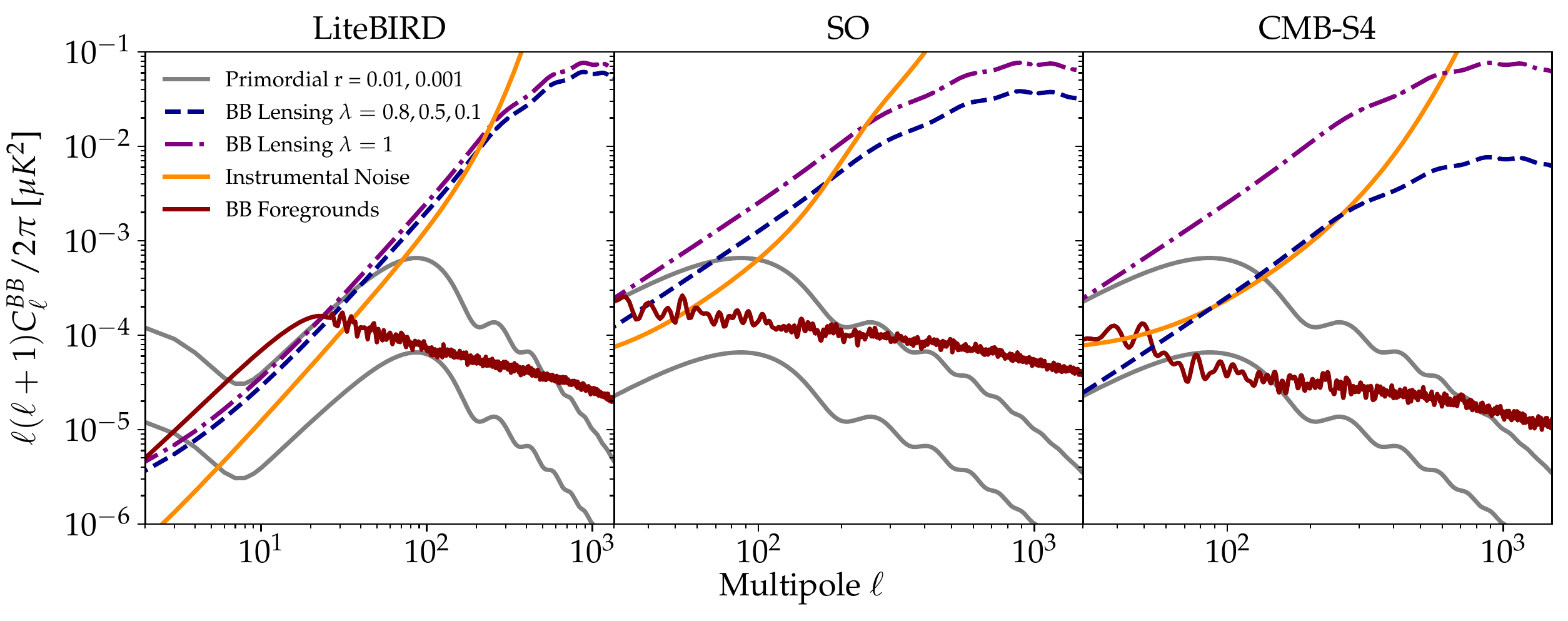}
    	\caption{The power spectrum of foregrounds residuals for dust and synchrotron  (solid red curve) and the post-component separation instrumental noise (solid orange curve) for the three experiments considered in this work, the lensing contribution $\lambda C_{\ell}^{XX',\,lens}$ with delensing factor  $\lambda = 0.8$ for LiteBIRD, $\lambda = 0.5$ for SO and $\lambda = 0.1$ for CMB-S4 (dashed blue curve) and without delensing ($\lambda=1$, dot-dashed purple curve). We also plot as a reference the primordial $BB$ power spectrum $C_{\ell}^{BB, prim}$ for $r = 0.01$ (upper solid grey curve) and $r=0.001$ (lower solid grey curve).\label{fig:foregrounds_spectra}}
    \end{figure}

    
\subsection{Uncertainties from foreground cleaning}\label{sec:foregrounds:contamination}
Several emission mechanisms contribute to the diffuse foregrounds from our own Galaxy \citep[see][and references therein]{Dickinson}. 
In our analysis we consider thermal dust and synchrotron radiation -- respectively emitted by thermal emission of dust grains and cosmic ray electrons spiraling the Galactic magnetic field. They are most important contaminants to the CMB B-modes from CGB. We leave to a future work the inclusion of other contaminants of secondary importance, like spinning and magnetic dust \citep{draine_lazarian, draine_hensley} and carbon monoxide \citep{greaves,puglisi_etal_2017}.

As we anticipated, we will exploit the publicly available code FGBuster which represents an implementation of parameter fitting in foreground estimation and removal for CMB experiments. We review here very briefly the corresponding formalism used for the computation of the uncertainties after component separation, which is based on the parametric maximum likelihood approach \citep{errard_stivoli_stompor_2011, cmb4cast, stompor_etal_2016, stompor_leach_stivoli_baccigalupi}, and is the basis of the FGBuster implementation. In the presence of multi-component emissions contributing to the signal measured on a given sky pixel $p$, we can write 
\begin{equation}\label{eq:linear_fgs}
	\textbf{d}_{p} = \textbf{A}\textbf{s}_{p} + \textbf{n}_{p},
\end{equation}
where the data vector $\textbf{d}_{p}$ contains the multi-frequency measurements for the sky pixel $p$, $\textbf{s}_{p}$ is the multi-component sky signal vector (with each polarized sky component represented by an entry of the vector), $\textbf{A}$ is the mixing matrix and $\textbf{n}_{p}$ is the instrumental noise vector. The instrumental noise at each frequency is assumed known a priori to be Gaussian and uncorrelated, with variance matrix $\textbf{N}_p$.
The columns of the mixing matrix are the SEDs of the components. They are not completely determined a priori and can have free parameters -- the so-called spectral parameters $\beta$ --
that have to be determined from the frequency maps.

In our analysis we consider three components: CMB, thermal dust and synchrotron. We assume perfect calibration and therefore the CMB emission law has no free parameters and is constant in thermodynamic units. For the frequency dependence of the synchrotron emission, we assume its brightness temperature to be a curved power-law 
\begin{equation}\label{eq:power_law_sync}
	\textbf{A}_{sync}(\nu) = \left(\frac{\nu}{\nu_{0}}\right)^{\beta_{s}+C_{s} \log(\nu/\nu_{0})}.
\end{equation}
where $\nu_{0}$ is the reference and pivot frequency of the synchrotron emission and is fixed at 70 GHz. The spectral index $\beta_{s}$ and the curvature $C_{s}$ are the free parameters. Note that, as of today, no evidence was found for the curvature \citep[see][and references therein]{Krachmalnicoff2018}. However, the experimental configurations that we consider will have a much higher sensitivity and therefore their results can be influenced by small departures from the standard power-law emission. For dust, we assume the standard modified black-body 
\begin{equation}\label{eq:power_law_dust}
	\textbf{A}_{dust} (\nu) = \left(\frac{\nu}{\nu_{0}}\right)^{\beta_{d}+1} \frac{e^{\frac{h\nu_{0}}{kT_{d}}}-1}{e^{\frac{h\nu}{kT_{d}}}-1},
\end{equation} 
where in this case $\nu_{0}$ is chosen equal to 353 GHz. $\beta_{d}$ is the spectral index of the emissivity and $T_d$ is the temperature of the grains, and they are both free parameters.
In total, we have four free parameters. Their reference values (i.e. the ``true'' values that we assume in the forecast) are $\beta_{s}= -3.0$, $C_{s}= 0$, $\beta_{d}= 1.54$ and $T_{d} = 20$ K, which well represent current constraints and measurements \citep{planck_2018_foregrounds}. 

The component separation process first estimates the non-linear parameters and then uses the estimated mixing matrix to separate CMB and foregrounds through a linear combination of the frequency maps. These two steps are shared by many component separation approaches, we refer to \cite{stompor_leach_stivoli_baccigalupi} for more details on the specific procedure that we consider. We can identify two contributions to the uncertainty of the estimated CMB map, both sourced by the instrumental noise. In the first step, the statistical uncertainty in the determination of the spectral parameters and the consequent imperfect estimation of the emission laws causes a leakage of foreground power into the CMB map -- the so-called \emph{statistical foreground residuals}. Even if the mixing matrix were perfectly recovered, in the second step the noise in the frequency maps propagates to the component maps and is referred to as \emph{statistical noise}. These uncertainty terms add extra power to the CMB map. We forecast these contributions following \citep{errard_stivoli_stompor_2011}.

In order to estimate the statistical foregrounds residuals we first evaluate the statistical uncertainty on the spectral parameters as the inverse of their Fisher information 
\begin{align}\label{eq:likelihood}
\boldsymbol{\Sigma}^{-1} &\simeq 
-\textrm{tr}\left\{\left[  \frac{\partial \textbf{A}^{T}}{\partial\beta} \textbf{N}^{-1} \textbf{A} (\textbf{A}^{T} \textbf{N}^{-1} \textbf{A})^{-1} \textbf{A}^{T} \textbf{N}^{-1} \frac{\partial \textbf{A}}{\partial\beta'} - \frac{\partial \textbf{A}^{T}}{\partial\beta} \textbf{N}^{-1} \frac{\partial \textbf{A}}{\partial\beta'}\right]\sum_{p} \textbf{s}_{p}\textbf{s}^{T}_{p}\right\}.
\end{align}
The power spectrum of the foreground residuals is equal to
\begin{equation}
    C_{\ell}^{fgs} \equiv \sum_{k,k'}\sum_{j,j'}\boldsymbol{\Sigma}_{kk'}\alpha_{k}^{0j}\alpha_{k'}^{0j'}C_{\ell}^{jj'},
\end{equation}
where 
\begin{equation}
    \alpha_{k}^{0j} \equiv - \left[  (\textbf{A}^{T} \textbf{N}^{-1} \textbf{A})^{-1} \textbf{A}^{T} \textbf{N}^{-1} \frac{\partial \textbf{A}}{\partial\beta_{k}} \right]_{0j}\,,
\end{equation}
and $j$ and $j'$ run over dust and synchrotron and 0 is the component-index of the CMB. 

This estimation assumes spatially-constant spectral parameters, which is probably a too stringent assumption given the high sensitivity of the experimental configurations we consider. Therefore we suppose to fit the spectral parameters independently over patches equal to HEALPix pixels with resolution parameter $N_{side}=8$, corresponding to an extension of about 7 degrees in the sky. The adopted value reflects the current knowledge concerning the typical angular scale of spatial variation of foreground spectral parameters \citep{planck_2018_foregrounds}, currently implemented in foreground models \citep{pysm_ref}. 
The number of patches is $n_{patch}=[12\times f_{sky} \times N_{side}^{2}]$ and, assuming statistical isotropy of the foregrounds, this factor rescales upwards ${\bf \Sigma}$ and, consequently, $C_{\ell}^{fgs}$. This estimation follows \cite{cmb4cast} and works on scales smaller than the patch size, 
$\ell_{patch}\simeq 25$. On larger scales the foregrounds residuals have the shape of a white spectrum because the noise and, consequently, the foregrounds residuals in each patch are uncorrelated \citep[see][for more details]{errard_stompor_2018}. For simplicity,
we take these large scales foreground residuals to be constant and equal to the statistical residuals around $\ell = \ell_{patch}$.

As far as the statistical noise is concerned, we follow \citep{stompor_etal_2016} and estimate it as
\begin{equation}
      C_{\ell}^{XX',\,noise} \equiv  \left[ \left(\textbf{A}^{T} \left(\textbf{N}_{\ell}^{XX'}\right)^{-1}\textbf{A}\right)^{-1}\right]_{CMB \, \,CMB}\, ,
\end{equation}
where  $\textbf{N}_{\ell}^{XX'} \equiv \left(N_{\ell}^{XX'}\right)^{\nu \nu'} \equiv N_{\ell, \nu}^{XX'} \, \delta^{\nu'}_{\nu}$ is a matrix containing the instrumental noise for each frequency channel as we anticipated in Eq.~\ref{eq:noise}. Note that the term $N_{\ell, \nu}^{XX'}$ is diagonal in $X,X'$, since the instrumental noise is uncorrelated in $T$, $E$ and $B$. 

The statistical residuals and the noise after component separation are output by FGBuster, which contains the forecast tool xForecast \citep{stompor_etal_2016} and relies on PySM \citep{pysm_ref} for the simulation of the foreground emission.

We report the foreground residuals and the noise after component separation for LiteBIRD, SO and CMB-S4 in Figure \ref{fig:foregrounds_spectra}.
Already at this stage, we can see that the inclusion of the foreground residuals is always relevant for scales larger than a degree and is particularly important at the scales accessible from space. 



\subsection{Impact of foreground residuals on PCA}\label{sec:foregrounds:impact}
In order to evaluate the impact of foreground residuals, we compare the uncertainties $\sigma_{r}$ on the $T/S$ ratio obtained from the Fisher matrix $F_{ab}$ for the cosmological parameters (Section \ref{sec:fishertheory}) with and without the addition of foregrounds. As we can see from the values reported in Table \ref{table:fgs_vs_nofgs} for the three considered experimental configurations and three values of $r$, a proper inclusion of the uncertainty coming from the foregrounds is most important for the LiteBIRD configuration, with an increase in $\sigma_{r}$ by about a factor $\sim 4$ (assuming $r=0$). The SO configuration is the least affected by foregrounds, being dominated by the instrumental noise contribution, with $\sigma_{r}$ growing by a factor between $1.25$ and $1.44$, depending on the value of $r$ we are assuming. CMB-S4 shows a definitely significant increase between, with the uncertainty on $r$ taking a factor ranging between $1.25$ and $2.3$. 

 \begin{table}
		\centering
	\footnotesize{
		\begin{tabular}{|c|l|c|c|c|}
			\hline \hline
			Experiment 
			& $r$
			& $\sigma_{r}$
			& $\sigma_{r, NOFG}$
			& Ratio
			\\
			\hline	\hline
			    & $r=0$ & $2.7\times 10^{-4}$ & $6.6\times 10^{-5}$ & 4.1  \\
			LiteBIRD & $r=0.001$ & $4.0\times 10^{-4}$ & $2.0\times 10^{-4}$ & 2.0 \\
			   & $r=0.01$& $6.0\times 10^{-4}$ & $4.4\times 10^{-4}$ & 1.36\\
			
			\hline
			   & $r=0$& $8.2\times 10^{-4}$ & $5.7\times 10^{-4}$ & 1.44 \\
			SO & $r=0.001$ & $8.6\times 10^{-4}$ & $6.2\times 10^{-4}$ & 1.39\\
			  & $r=0.01$ & $1.3\times 10^{-3}$ & $1.04\times 10^{-3}$ & 1.25\\
			
			\hline
			  & $r=0$& $4.6\times 10^{-4}$ & $2.0\times 10^{-4}$ & 2.3\\
			CMB-S4 & $r=0.001$ & $5.4\times 10^{-4}$ & $3.0\times 10^{-4}$ & 1.8\\
			   & $r=0.01$ & $1.2\times 10^{-3}$ & $9.6\times 10^{-4}$ & 1.25\\
				\hline\hline
		\end{tabular}}
	\caption{Comparison between uncertainty on $T/S$ ratio $r$ from Fisher matrix including ($\sigma_{r}$) or not including ($\sigma_{r, NOFG}$) foregrounds residuals into the analysis, for three experimental configurations and three values of $r$. The ratio of the uncertainties on $r$ before and after the inclusion of foregrounds is also shown.}
	\label{table:fgs_vs_nofgs}
\end{table}

On the basis of these evidences, we proceed by addressing the variation in the properties of PCA modes with and without foregrounds. We apply the formalism developed in Section \ref{sec:PCA}} for deriving the PCA modes $S_{a}(k)$ and we restrict here to the LiteBIRD case with $r=0$. The results are shown in Figure \ref{fig:fg_vs_nofg}. For reasons which will be explained in the next Section, the $S_{a}(k)$ are obtained in this case by orthogonalizing with respect to all cosmological parameters except $r$. Consequently, the non-oscillatory modes, typically among the first ones, catch the overall power of tensor modes and play the effective role of the parameter $r$ itself.   

\begin{figure}
   	\centering
   	\includegraphics[scale=1.0]{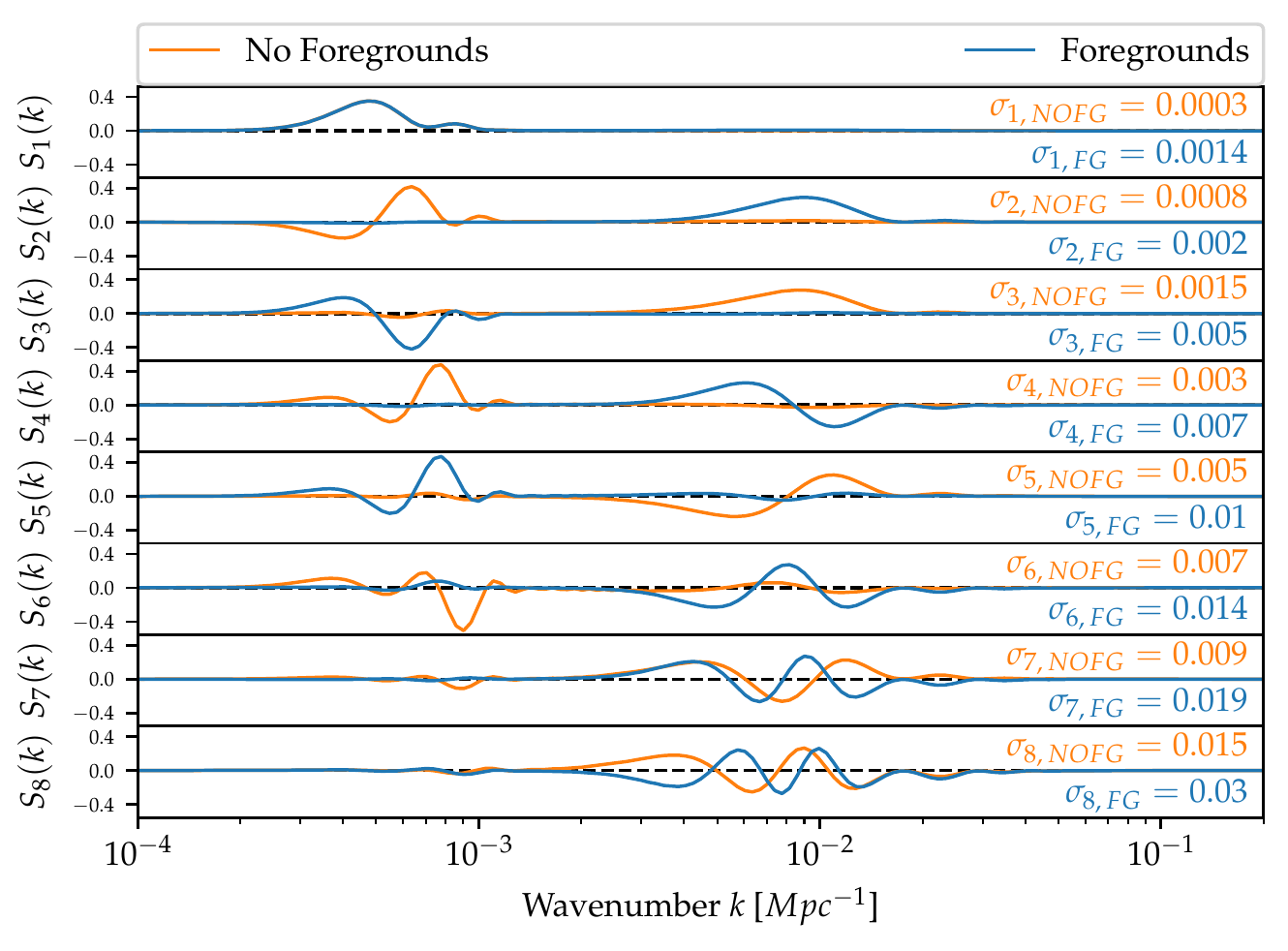}
   	\caption{A comparison between the first eight PCA modes, obtained with LiteBIRD specifications, assuming $r=0$ and delensing factor $\lambda = 0.8$, with (solid blue curve) and without (solid orange curve) the foregrounds residuals. The errors on the modes with (indicated as $\sigma_{a, \,FG}$ in blue) and without ($\sigma_{a, \,NOFG}$ in orange) foregrounds are also reported. Note that our results will based on the modes \textit{with} foregrounds.\label{fig:fg_vs_nofg}}
   \end{figure}

Both with and without foregrounds, we see two different positive modes that pick up power around $k\approx 6\times 10^{-4} {\rm Mpc}^{-1}$ and $k\approx 6\times 10^{-3} {\rm Mpc}^{-1}$, respectively. These are the scales contributing to the reionization and recombination bump \citep{hiramatsu_etal_2018}. The other modes present oscillatory patterns with characteristic oscillation scales getting shorter and shorter as $n$ increases, meaning that the experimental setup provides weaker constraints on smaller features of the primordial power spectrum.
The presence of foregrounds has visible and non-negligible effects on all the PCA modes and their eigenvalues. While the positive mode related to the reionization bump is the best constrained mode both with and without foregrounds, the one related to the recombination bump is upgraded from third to second mode when foregrounds residuals are considered, overtaking the first oscillatory pattern around $k\approx 6 \times 10^{-4} {\rm Mpc}^{-1}$. Note that the constraints on both modes are degraded, but the effect is more important for the latter since foreground residuals are the dominant uncertainty for the reionization peak, while they are lower than lensing and noise for the recombination peak. 
A closer look at the errors associated to each mode (also reported in Figure \ref{fig:fg_vs_nofg}) shows how foregrounds lead to a dramatic loss of information on the reionization bump. This information is indeed carried by the foregrounds-free modes $\mathcal{S}_{1,\,NOFG}$ and $\mathcal{S}_{2,\,NOFG}$ -- which corresponds to the modes $\mathcal{S}_{1,\,FG}$ and $\mathcal{S}_{3,\,FG}$. Their uncertainty grows by a factor $\sim 5$ (from $\sigma_{1,\, NOFG}=0.0003$ to $\sigma_{1, \,FG}=0.0014$) and by a factor $\sim 6$ (from $\sigma_{2,\,NOFG}=0.0008$ to $\sigma_{3, \,FG}=0.005$). In comparison, the information loss on the recombination bump is restrained: the constraint on the foreground-free mode $\mathcal{S}_{3,\, NOFG}$ -- which corresponds to $\mathcal{S}_{2,\, FG}$ and carries most of the information on the recombination bump -- is just $\sim 30\%$ smaller ($\sigma_{3,\, NOFG}=0.0015$ instead of $\sigma_{2,\, FG}=0.002$). Note that shifts of relative importance when foregrounds are introduced are also present in the other modes (see Figure \ref{table:fgs_vs_nofgs}) and, in all the cases, the uncertainties $\sigma_{a}$ associated to each PCA mode $S_{a}$ increase.  
The SO and CMB-S4 configurations do not get constraints from the reionization bump, and therefore the shape and relative importance of the PCA modes do not change when foregrounds are introduced. However, as in the LiteBIRD case, their constraints on all the modes are reduced. In the CMB-S4 case the effect of foregrounds is certainly non-negligible, with the uncertainties on the first two modes taking a factor $2.5$ and $2$, respectively. Concerning SO, since the instrumental noise is higher, the presence of foreground residuals is less important but still noticeable, with enhancements by a factor $1.36$ on $\mathcal{S}_{1}$ and $1.18$ on $\mathcal{S}_{2}$.
 

We conclude that the contribution of the foreground residuals definitely cannot be neglected in a PCA analysis of the  primordial tensor perturbations in the context of $B$-mode experiments. The inclusion of foregrounds will therefore be the baseline for our analysis in the prosecution of this work.

\section{Constraining primordial tensor perturbations with PCA}\label{sec:results}
We now discuss the application of the PCA formalism described in Sections \ref{sec:PCA} and \ref{sec:foregrounds} to the experimental configurations we consider.  
As emphasized before, the application to the tensor power spectrum needs some special cares compared to the scalar power spectrum case. The reason is that in the former case the PCA basis does not describe small deviations of a large, well constrained power spectrum. 
One of the most important consequences is that, if some power from primordial gravitational waves is found in the B-modes power spectrum, the relative constraints on the PCA amplitudes can be very different from the ones predicted by the PCA analysis itself (i.e. $\sigma_i < \sigma_j$ for $i < j$).  
In other words, the information in a power spectrum with a power contribution from tensors can be very different from the information matrix that defined the PCA basis. This is potentially problematic because the extent of the tensor contribution is not known a priori but the number of modes retained is fixed to $N$ from the onset, discarding the information lying outside the space spanned by $\{\mathcal{S}_a \mathcal{S}_a^T\}_{a < N}$. In the scalar case we just have to choose a sufficiently high $N$ to capture most of the Fisher information from which the $\mathcal{S}_a$ themselves where defined. In the tensor case we have to test the stability of this property with respect to a range of possible Fisher information matrices $\mathbf{F}_r$, arising from angular power spectra different from the one used for the definition of the $\mathcal{S}_a$ basis (which assumes no tensors). 
We can think of the sum of the eigenvalues of $\mathbf{F_r}$ as the total information about the primordial tensor power spectrum. It can be computed as $tr(\mathbf{F_r})$, without performing a Singular Value Decomposition (SVD). The Fisher information on the first $N$ $m_a$ coefficients, on the contrary, is equal to $\mathbf{S}_N^{T}\, \mathbf{F}_{r}\, \mathbf{S}_N$, where the matrix $\mathbf{S}_N$ contains the first $N$ columns of $\mathbf{S}$. Therefore, when we describe the power spectrum only in terms of the first $N$ $\mathcal{S}_a$ modes, the fraction of the total information we retain is given by
\begin{equation}\label{eq:information_fraction}
I(r, N) =\frac{tr \left(\mathbf{S}_N^{T}\, \mathbf{F}_{r}\, \mathbf{S}_N\right)}{tr \left(\mathbf{F}_{r}\right)}.
\end{equation}
For $r=0$, our parametrization of the tensor power spectrum is by construction the one that guarantees the maximum $I$ for any value of $N$. For $r \neq 0$ this is no longer true, but a value of $I$ sufficiently high means that the parametrization still capture (approximately) all of the information that our data can provide.
We choose $N$ such that $I$ is high enough ($98\%$) up to $r < 0.01$. Once $N$ is fixed, the Fisher uncertainties associated with the $i$-th PCA mode in our basis are given by
\begin{equation}\label{eq:inversion}
    \sigma^{2}_{i} = \left(\mathbf{S}_{N}^{T}\, \mathbf{F}_{r}\, \mathbf{S}_{N}\right)^{-1}_{ii}\,,
\end{equation}
Note that for $r \neq 0$ it is no longer guaranteed that $\sigma_i < \sigma_j$ for $i < j$ and that $m_i$ and $m_j$ are uncorrelated for $i \neq j$.
After studying the $I$ function and choosing the appropriate value of $N$, we study the PCA products for three observed CMB power spectra, computed for a scale invariant primordial power spectrum with $r = 0$, $0.01$ and $0.001$.

\subsection{Application to LiteBIRD}\label{sec:litebird}
\begin{figure}
	\centering
		\includegraphics[scale=0.65]{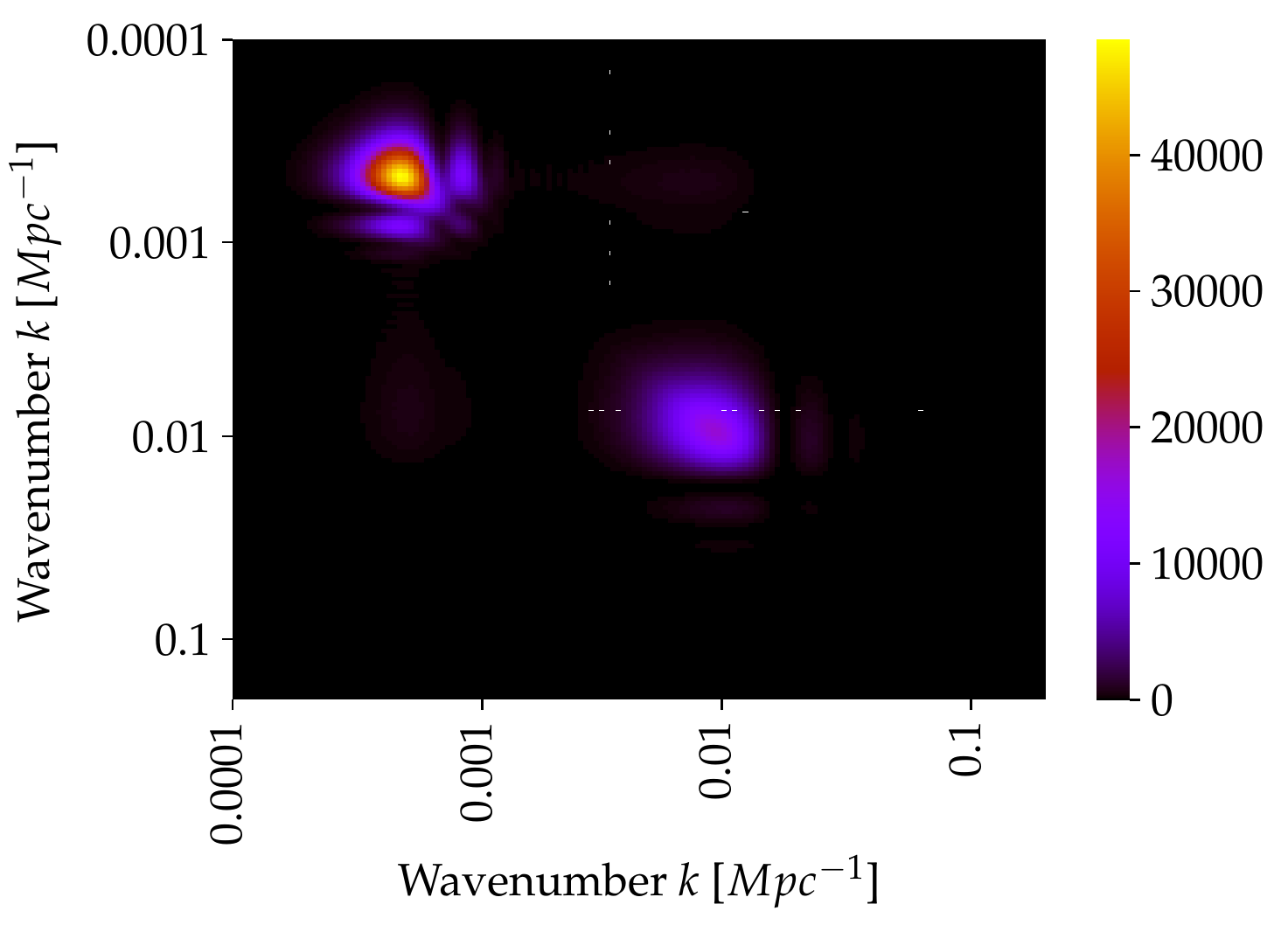}
		\label{fig:1}
\caption{Fisher matrix for LiteBIRD, for $r=0$ and delensing factor $\lambda=0.8$. The color bar gives the value of the elements of the matrix, with higher value representing more information. \label{fig:fisher_litebird}}  
\end{figure}

The Fisher matrix for $r=0$, which we use to define the PCA basis, is shown in Figure \ref{fig:fisher_litebird} and its first eigenvectors are the blue functions reported in Figure \ref{fig:fg_vs_nofg}. We can recognize in the matrix the features that we highlighted in Section \ref{sec:foregrounds:impact}, when discussing the functions. Most of the information is concentrated in two regions, precisely the scales that contribute the most to the recombination and the reionization bumps, in accord with \citep{hiramatsu_etal_2018}. The strong off-diagonal correlations express the fact that we are chiefly sensitive to the overall power in those regions, while we have much less information about features within them. 

\begin{figure}
	\centering
	\includegraphics[scale=0.7]{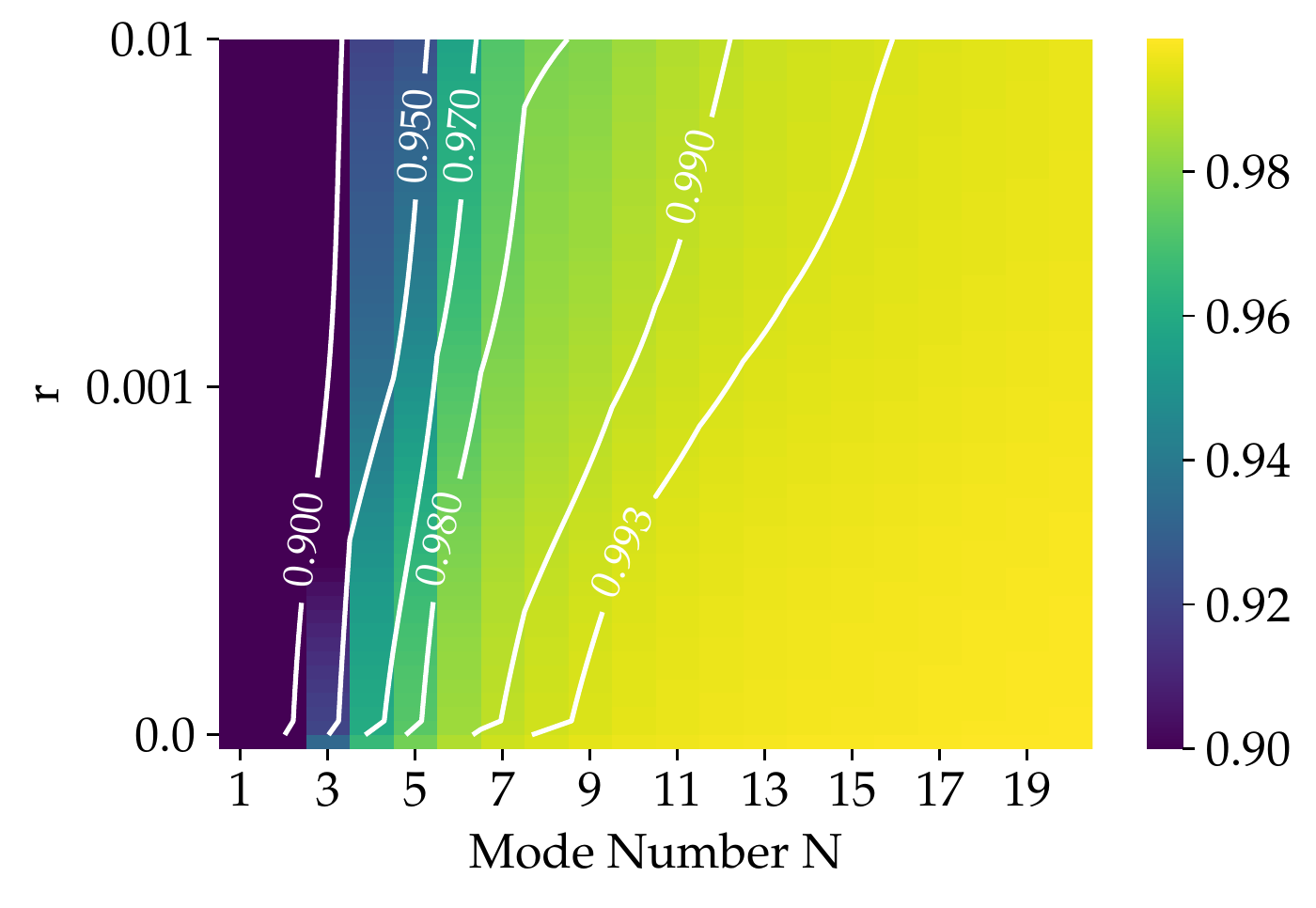}
	\caption{Information fraction $I(r,N)$ (Eq.~\ref{eq:information_fraction}) as a function of $r$ and number of modes retained in the analysis $N$, using the LiteBIRD experimental configuration. The level curves of constant $I(r,N)$ for 90\%, 95\%, 97\%, 99\% and 99.3\% are reported in white. To obtain this plot we generated Fisher matrices $\mathbf{F}_r$ for 50 logarithmically spaced values of $r$ between $0.0001$ and $0.01$ plus the value $r=0$. \label{fig:information_LiteBIRD}}
\end{figure}

Figure \ref{fig:information_LiteBIRD} shows the evaluation of $I$ over a grid of values of $r$ and $N$. Considering that $r$ axis is log-scaled, the level curves are quite vertical, meaning that the PCA basis defined for $r= 0$ is effective in capturing the information in an observation with a very different primordial power spectrum. While for $r = 0$ we need five modes in order to capture 98\% of the information, we only need three more modes when $r = 0.01$. For LiteBIRD we indeed set N = 8.

In Table \ref{table:fisher_pure} we compare the uncertainties and the signal-to-noise ratio $S/N=(m_{a}^{2}/\sigma_{a}^{2})^{1/2}$ associated to each of the modes retained for three values of $r$ . We can see from this table that the values of $\sigma_{a}$ for the case $r=0$ are monotonically increasing -- as expected -- but those of the $r \neq 0$ cases are not. In particular, in these latter cases the second mode has smaller uncertainty than the first one. Reminding the shape of these modes, we see that this means that as $r$ is different from zero, we get more information from the recombination bump than from the reionization bump. This trend is also reflected in the associated $S/N$ ratio, which is greater in the second mode than in the first for both $r=0.001$ and $r=0.01$. Moreover, the $S/N$ ratio suggests that only the second mode, for both $r\neq0$ cases, can be detected.
However, one should not interpret this as a limit of PCA, but as a product of the fact that we are performing PCA over a signal simulated with a flat tensor power spectrum characterized only by its $T/S$ ratio, while PCA is designed to describe power spectra with more features. We refer to Section \ref{sec:earlyuniversemodels} for a toy example of early universe model where the application of the PCA is more suited. 

\begin{table}
		\centering

		\footnotesize{	\begin{tabular}{|c|c|c|c c|c c|}
			\hline \hline
			 \multirow{2}{*}{Experiment} &
			 \multirow{2}{*}{PCA mode} &
			 \multirow{2}{*}{$r=0$} &
			\multicolumn{2}{|c|}{$r=0.001$ } & 
			\multicolumn{2}{|c|}{$r=0.01$ } \\
			\cline{4-7}
			&
			& 
			& $\sigma_{a}$
			& $S/N$
			& $\sigma_{a}$
			& $S/N$
			\\
			\hline	\hline
			         & 1st & 0.0014 &  0.005 & 1.0 & 0.03 & 2.0 \\
			         & 2nd & 0.002  & 0.002 & 3.0 & 0.003 & 16.0\\
			         & 3rd & 0.005  &  0.01 & 0.09 & 0.06  & 0.14\\
  \textbf{LiteBIRD} & 4th & 0.007  &  0.007 & 0.13 & 0.009 & 1.02\\
			         & 5th & 0.01   &  0.019 & 0.14 & 0.09  & 0.3 \\
			         & 6th & 0.014  &  0.014 & 0.17 & 0.02 & 1.2 \\
			         & 7th & 0.019  &  0.019 & 0.02 & 0.02 & 0.2 \\
			         & 8th & 0.03   &  0.03 & 0.08 & 0.04 & 0.6 \\

			\hline\hline
			         & 1st & 0.004 & 0.004 & 1.2 & 0.006 & 8.0  \\
		 \textbf{SO} & 2nd & 0.01 & 0.01 & 0.08 & 0.014 & 0.6 \\
			         & 3rd & 0.019 & 0.019 & 0.14 & 0.03 & 0.9 \\
	                 & 4th & 0.03 & 0.03 & 0.006 & 0.04 & 0.05 \\
			\hline\hline
			         & 1st & 0.002 & 0.002 & 2.2 & 0.005 & 9.0\\
			         & 2nd & 0.006 & 0.006 & 0.09 & 0.014 & 0.4 \\
	\textbf{CMB-S4}  & 3rd & 0.01 & 0.01 & 0.3 & 0.02 & 1.4 \\
	                 & 4th & 0.014 & 0.019 & 0.005 & 0.03 & 0.03 \\
			         & 5th & 0.02 & 0.02 & 0.14  & 0.04 & 0.7 \\
			         & 6th & 0.03 & 0.03 & 0.04 & 0.05 & 0.2 \\
\hline\hline
		\end{tabular}}

	\caption{Comparison between the uncertainties from Fisher analysis for the three values of $r$ and the three experiments considered in this work. The signal-to-noise ratio ($S/N$) for the two cases $r=0.001$ and $r=0.01$  is also reported. 
	}
	\label{table:fisher_pure}
\end{table}

Even if we do not report the results in detail, we have also run the case with no delensing ($\lambda=1.0$ instead of $\lambda=0.8$) and, as expected, we see that delensing also contributes in shifting the relative importance in terms of information, from the reionization to the recombination bump.

\subsection{Application to SO and CMB-S4}\label{sec:SOandS4}

\begin{figure}
    \begin{subfigure}[b]{0.5\textwidth}
        \includegraphics[width=1.0\textwidth]{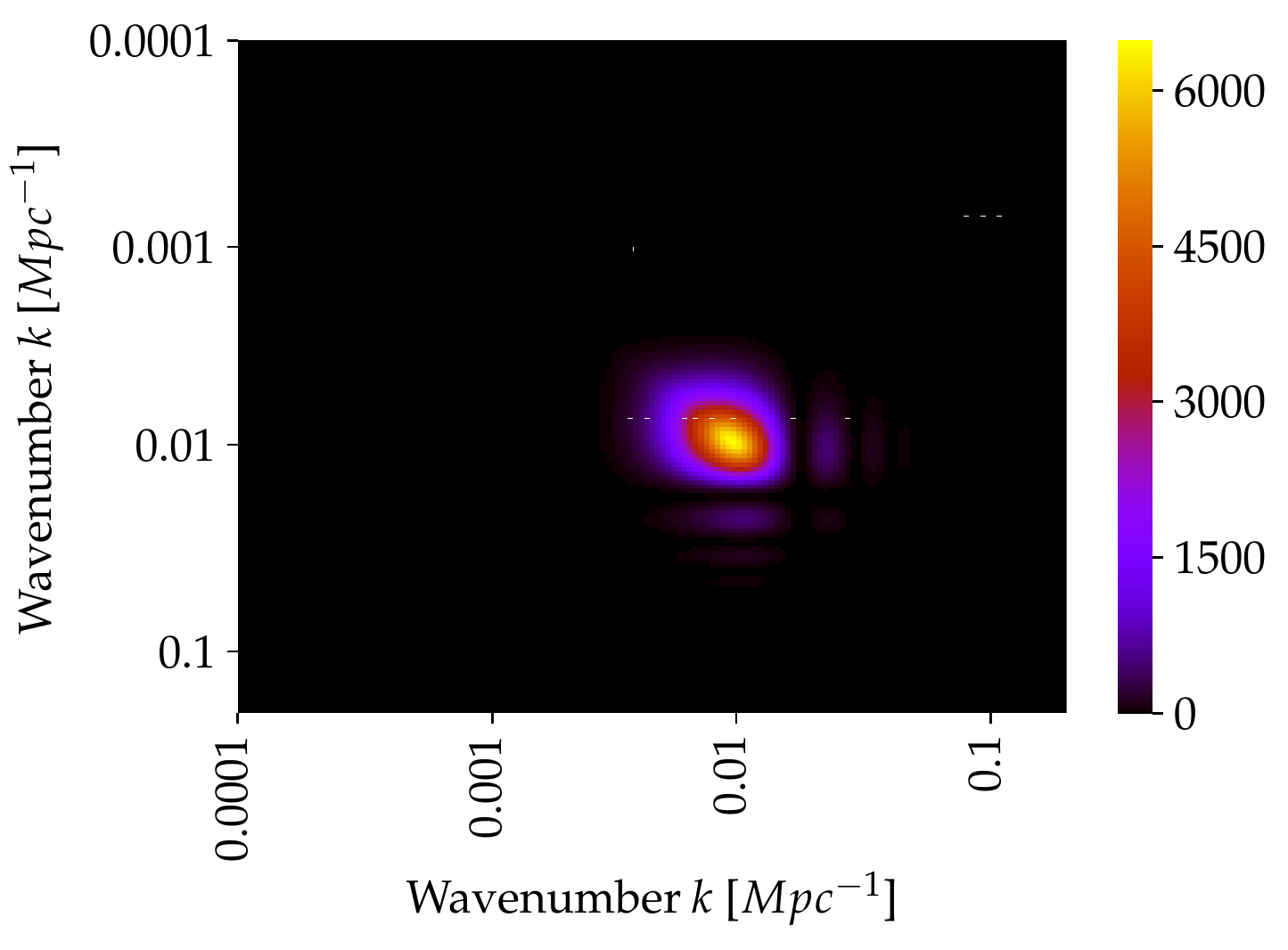} 
    \end{subfigure}
    \begin{subfigure}[b]{0.5\textwidth}
        \includegraphics[width=1.0\textwidth]{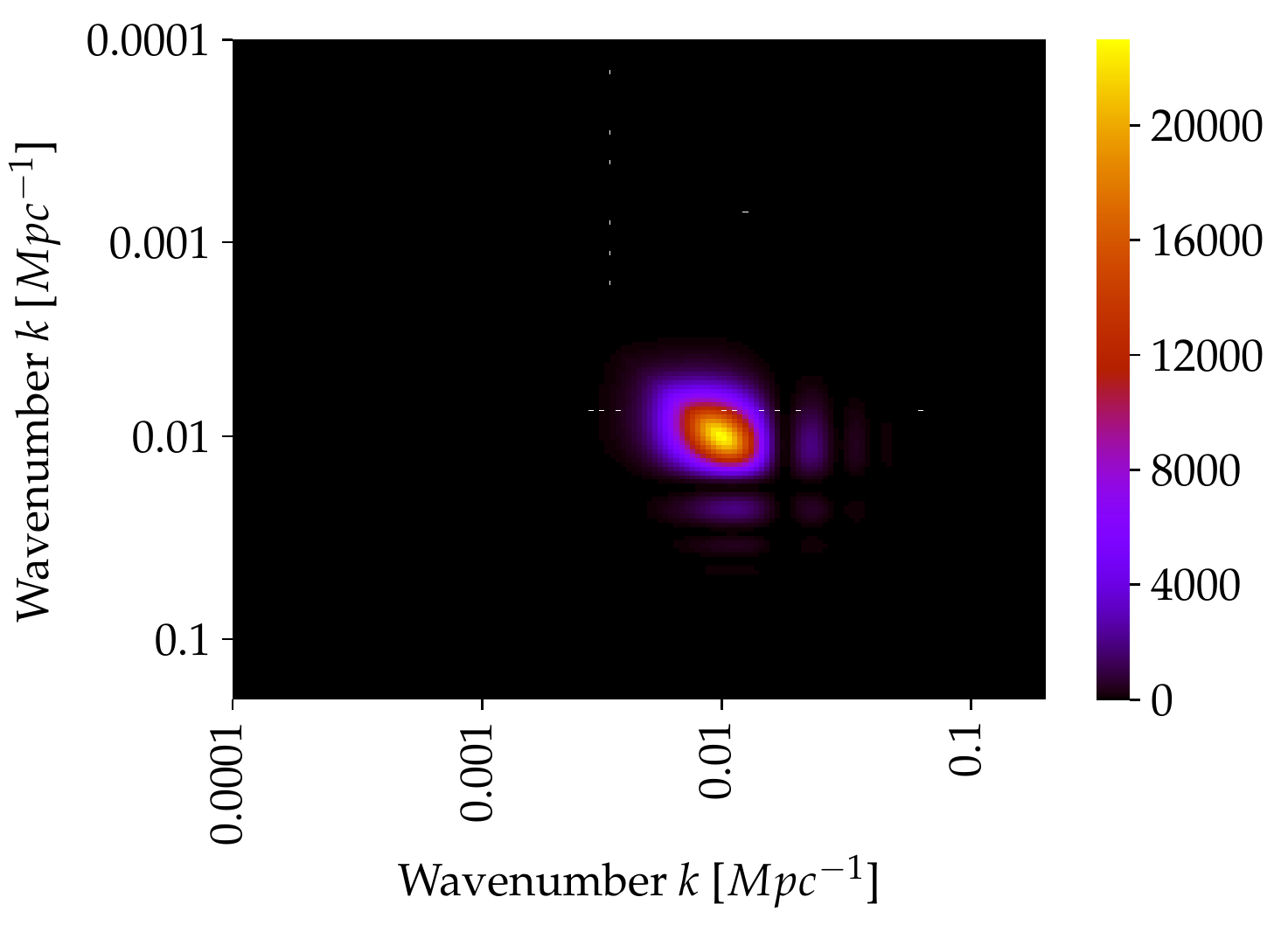} 
    \end{subfigure}
\caption{Fisher matrix for SO (left panel) and CMB-S4 (right panel) with $r=0$. The color bar gives the  value of the elements of the matrix. \label{fig:fisher_so_cmbs4}}
\end{figure}

\begin{figure}
	\centering
	\includegraphics[scale=0.8]{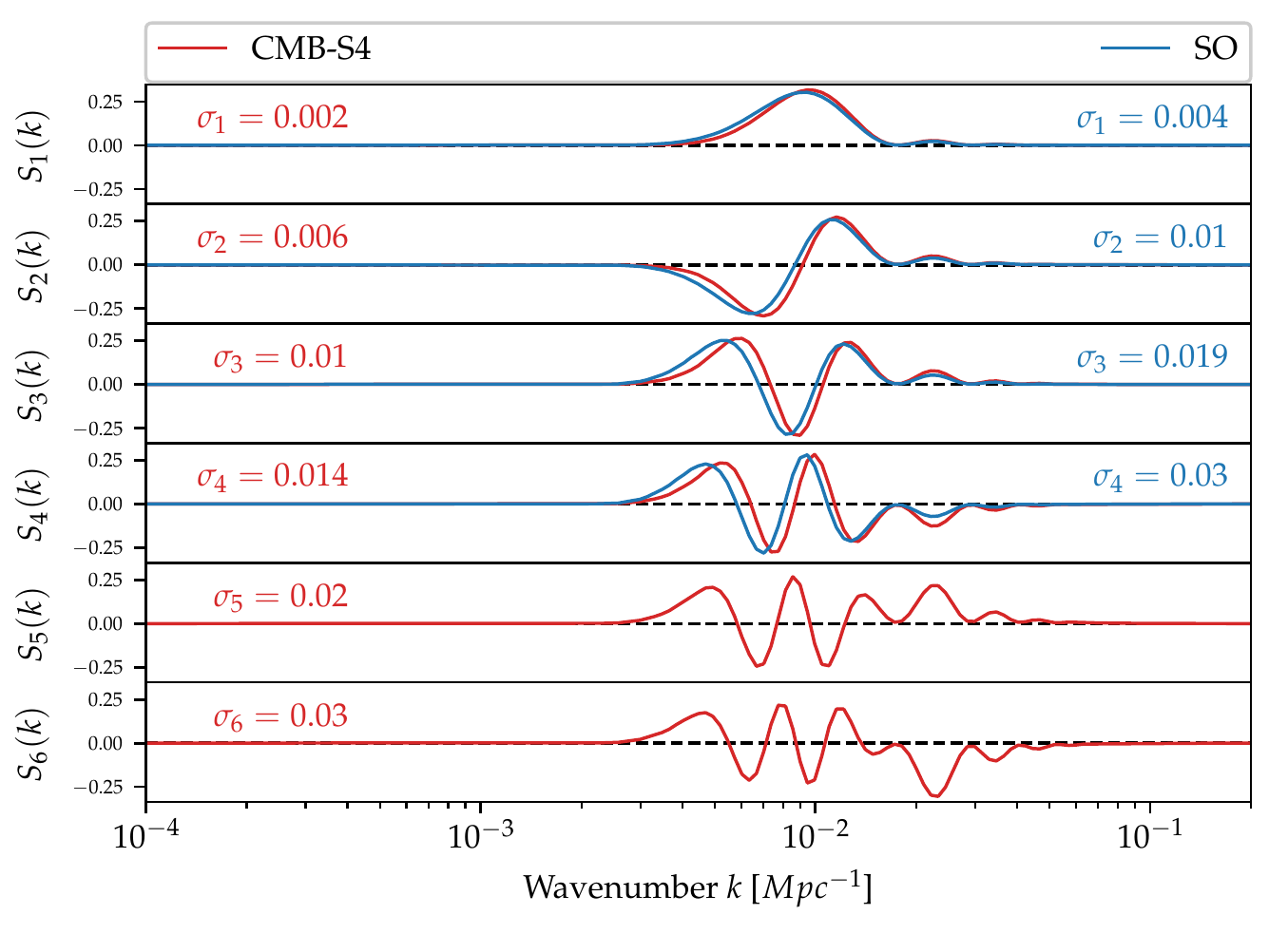}
	\caption{First four PCA modes for the SO (solid blue line) and first six for the CMB-S4 (solid red line) experiments with $r=0$. We also report the uncertainties on each mode for CMB-S4 (on the left) and for SO (on the right).  \label{fig:pca_so_baseline} 
	}
\end{figure}

We now focus on the two ground based experiments, SO and CMB-S4. We consider them together because the results are similar, as we shall see. 
To start with, the Fisher matrices for the two configurations (shown in Figure \ref{fig:fisher_so_cmbs4}) have essentially the same structure. By comparing with the LiteBIRD case (Figure \ref{fig:fisher_litebird}) we observe that SO and CMB-S4 are not sensitive to the scales that could be probed with the reionization bump. On the other hand, the higher resolution and delensing capability give access to some information beyond the first acoustic peak.
These features are visible also in the PCA modes, as we can see in Figure \ref{fig:pca_so_baseline}.

\begin{figure}
    \begin{subfigure}[b]{0.5\textwidth}
        \includegraphics[width=1.0\textwidth]{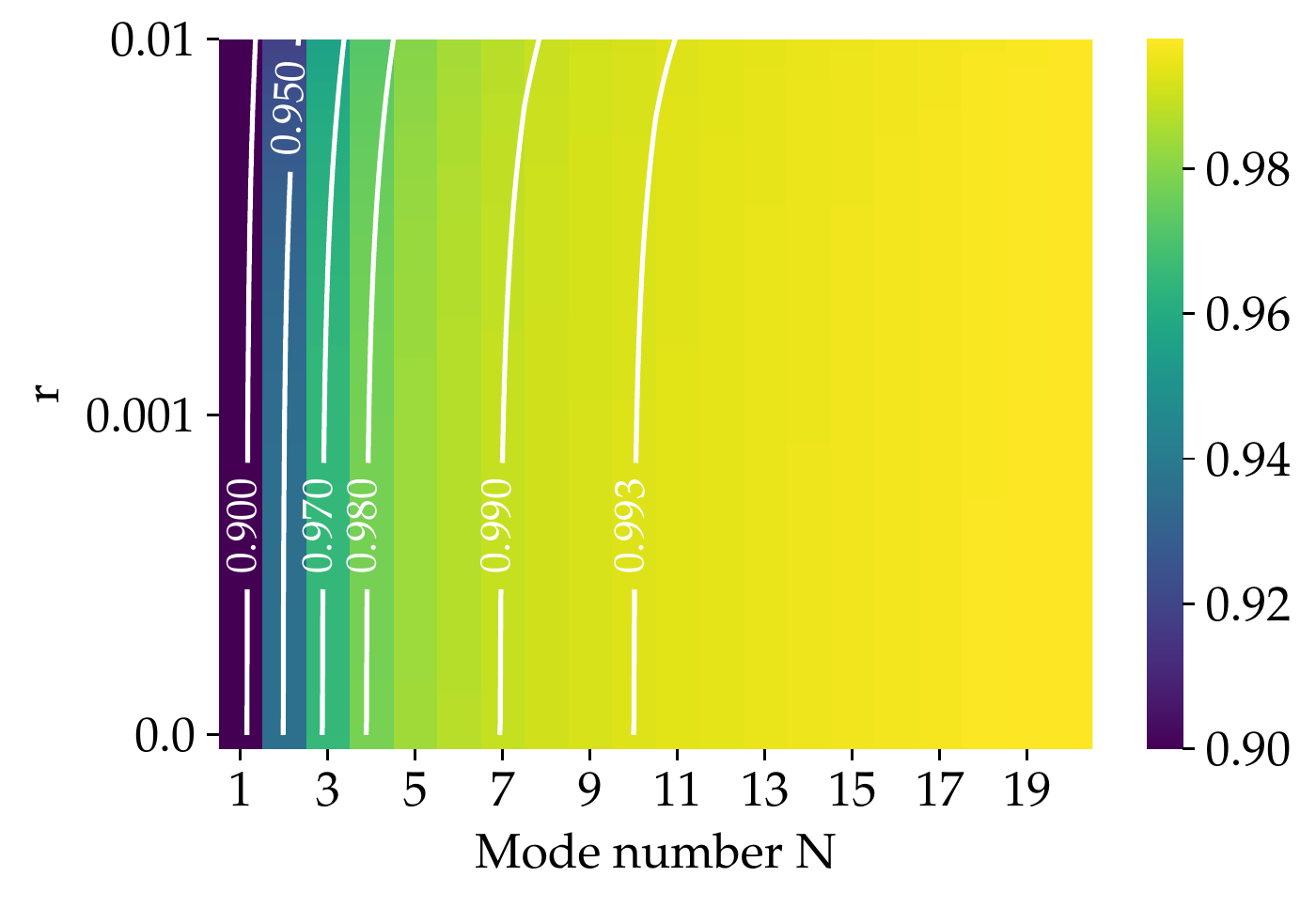} 
    \end{subfigure}
    \begin{subfigure}[b]{0.5\textwidth}
        \includegraphics[width=1.0\textwidth]{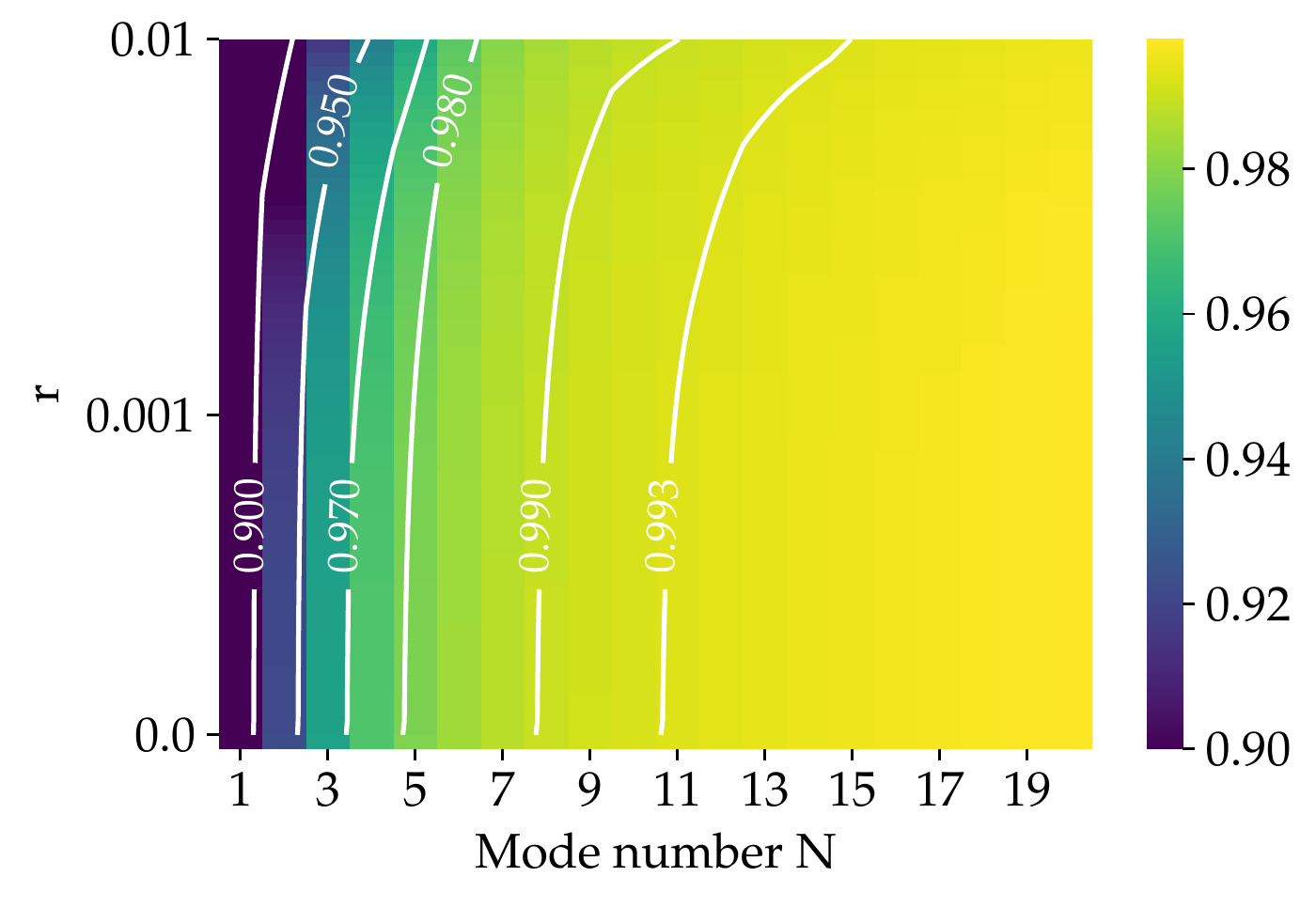} 
    \end{subfigure}
	\caption{Information fraction $I(r,N)$ (Eq.~\ref{eq:information_fraction}) as a function of $r$ and number of modes retained in the analysis $N$, using the SO (left panel) and the CMB-S4 (right panel) experimental configurations.
\label{fig:information_CMBS4}}
\end{figure}

In order to choose $N$ and make sure that we can use our PCA basis also for observed spectra that have power in the primordial tensor power spectrum, we study $I(r, N)$. This check is passed, as Figure \ref{fig:information_CMBS4} shows that the dependence on $r$ is even weaker than in the LiteBIRD case. We therefore need fewer modes to capture 98\% of the information in the whole $r$ range and use $N = 6$ for CMB-S4 and $N=4$ for SO. Note that the two configurations have similar $I$ for $r=0$ but differ for higher values of $r$, with the SO case being essentially $r$-independent. The reason is that SO has a higher noise than CMB-S4 and consequently it needs a higher signal with respect to the latter experiment to produce changes in the Fisher matrix.

The uncertainties on the PCA modes obtained for both experiments and the three different values of $r$ are reported in Table \ref{table:fisher_pure}, together with the associated $S/N$ ratio. 
Thanks to the higher sensitivity, the uncertainties of CMB-S4 are essentially always half of those SO. Beyond this difference they share the same qualitative behaviour. Notably, their uncertainty in the  $r=0.001$ case essentially coincide with $r=0$.

\subsection{Example of application: tilted spectra from inflation}\label{sec:earlyuniversemodels}
As we discussed in Section \ref{sec:mcmc}, PCA can be used to detect deviation of the tensor power spectrum with respect to the fiducial case.
In this section we show a very basic application of the PCA method, by applying it to data simulated using a red-tilted model of the tensor power spectrum.

\begin{figure}
    \begin{subfigure}[b]{0.47\textwidth}
        \includegraphics[width=1.0\textwidth]{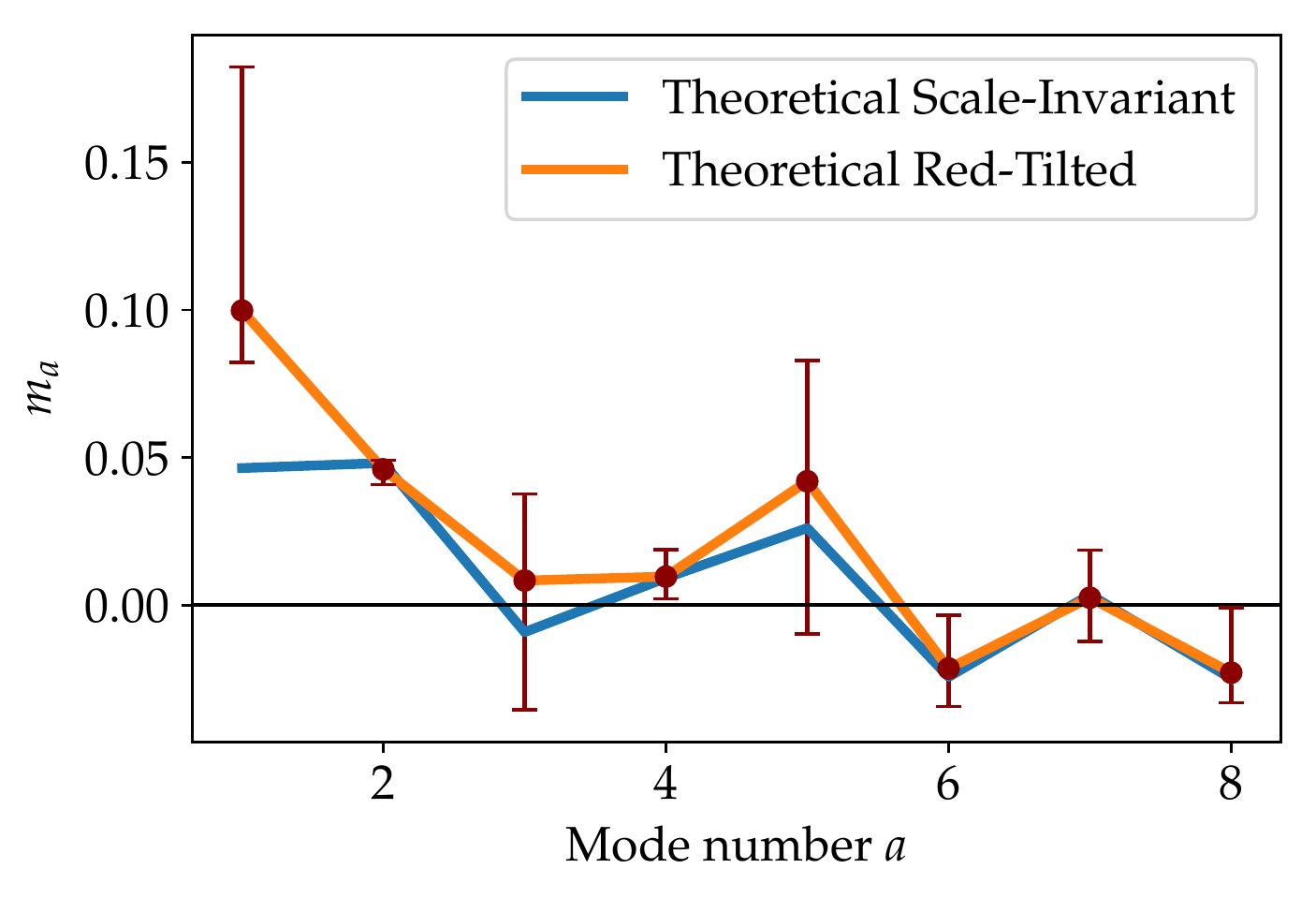} 
    \end{subfigure}
    \begin{subfigure}[b]{0.47\textwidth}
        \includegraphics[width=1.0\textwidth]{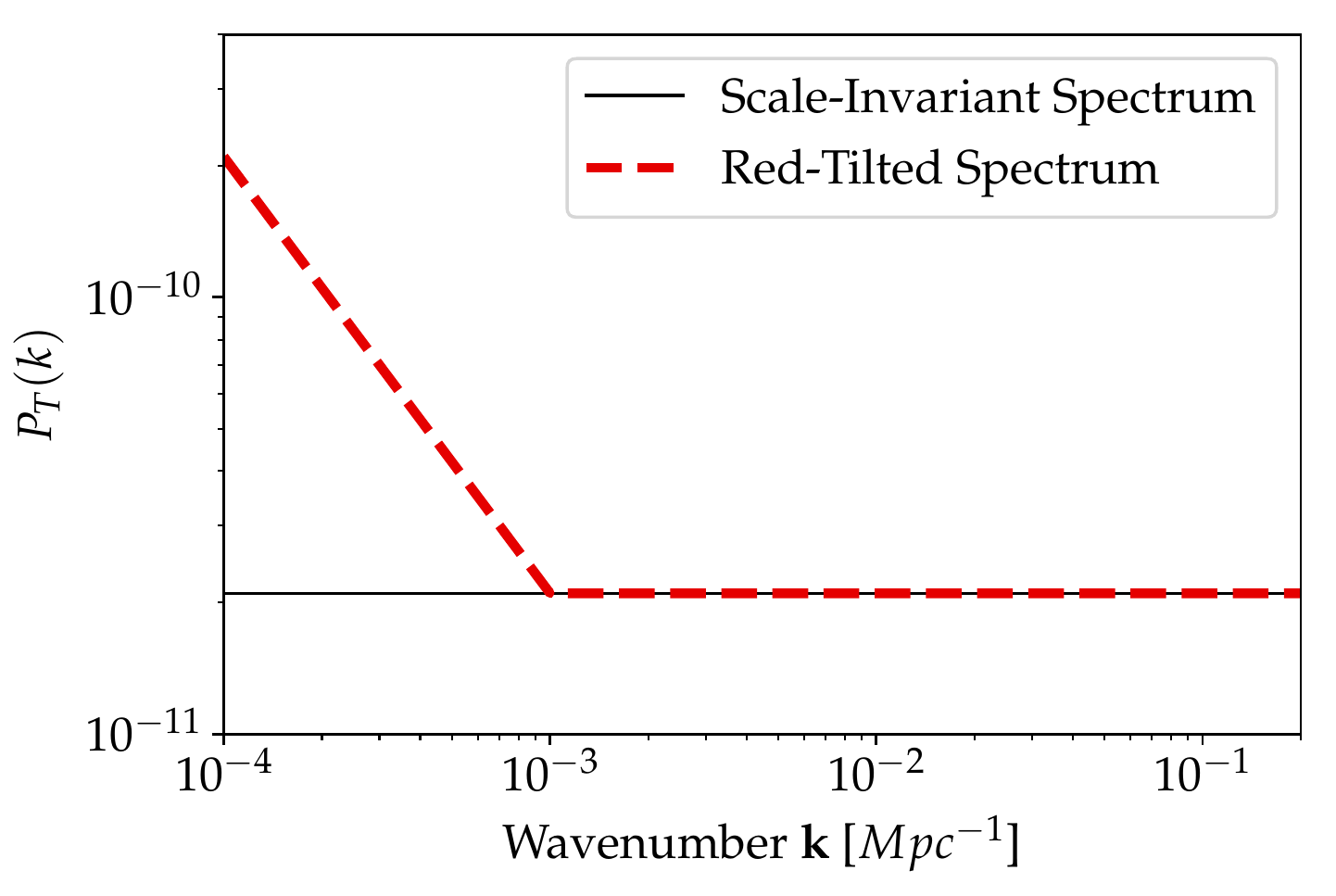} 
    \end{subfigure}
    \caption{Left panel: the red points represent the PCA amplitudes $m_a$ obtained from the MCMC for the red-tilted model as a function of the mode number $a$, with error bars obtained from the MCMC. In orange, the theoretical red-tilted tensor power spectrum model projected onto the PCA basis through formula \eqref{eq:projection}, in blue the theoretical scale-invariant power spectrum obtained via the same formula. Right panel: the red-tilted tensor power spectrum described in Section \ref{sec:earlyuniversemodels} (dashed red curve) and the scale-invariant power spectrum $A_{T}$ for $r=0.01$ (solid black line).\label{fig:redtilted}}
\end{figure}

We consider a model parametrized as ${\mathcal P}_{T}(k)= A_{T}
(k/k_0)^{n_{T1}}$ on scales $k<k_{1}$ and ${\mathcal P}_{T}(k)= A_{T}$ on scales $k \geq k_{1}$, with $(k_1=0.001\ {\rm Mpc}^{-1}, n_{T1}=-1)$ (right panel of Figure \ref{fig:redtilted}). This simple model has been considered also in \citep{hiramatsu_etal_2018} and \citep{farhang_sadr_2018}, because of its resemblance with the tensor power spectrum predicted by open inflation associated with bubble nucleation \citep{yamauchi_etal_2011}. However, we warn the reader that our choice of the model and the parameters $(k_1, n_{T1})$ is not motivated by fundamental physics and is only an example of exploitation of the PCA.  We leave the study of tensor power spectra from specific early universe physics to a future work. In this example we construct the PCA modes assuming the fiducial cosmology with $r=0$ and the specifications of the LiteBIRD satellite, while the data used as input for the MCMC are simulated assuming the red-tilted tensor power spectrum described above with $r=0.01$. The PCA amplitudes recovered from the MCMC for this red-tilted model are reported in the left panel of Figure \ref{fig:redtilted}, and are clearly consistent with projected ones (solid orange curve), showing a characteristic trend not compatible with the fiducial scale-invariant spectrum (solid blue curve). Also the cosmological parameters are very well recovered and compatible with the input ones, therefore this application represents a success for PCA.
We also compute a chi-square from the ratio of likelihoods for this red-tilted model and the fiducial scale-invariant model with $r=0.01$, obtaining $\chi^{2}\sim 8$. This value can be compared to the value $\chi^{2}\sim 20$ reported in Table 3 of \citep{hiramatsu_etal_2018} for the same power spectrum model and a similar experimental configuration which, however, does not include foregrounds. The difference between the two values is indeed due to the addition of foregrounds, since the red-tilted model starts to differ from a scale-invariant model with $r=0.01$ around the reionization bump scales, and, as we explained in Section \ref{sec:foregrounds:impact}, the constraints on these scales are strongly affected by the presence of foregrounds.

\subsection{Limitations of the PCA and MCMCs}\label{sec:limitations}
The PCA methodology we used to forecast the constraints on $\mathcal{P}_T$ is insensitive to the physicality prior $\mathcal{P}_T > 0$. In this section we discuss in detail the importance of such a prior and how it can jeopardize the validity of these Fisher estimates. While doing that, we also justify our choice of parametrizing $\mathcal{P}_T$ only as a linear combination of PCA modes, without including the popular cosmological parameter $r$. We indeed show that this parametrization makes the Fisher analysis marginally more robust against the physicality prior.

\begin{figure}
	\centering
	\includegraphics[scale=0.85]{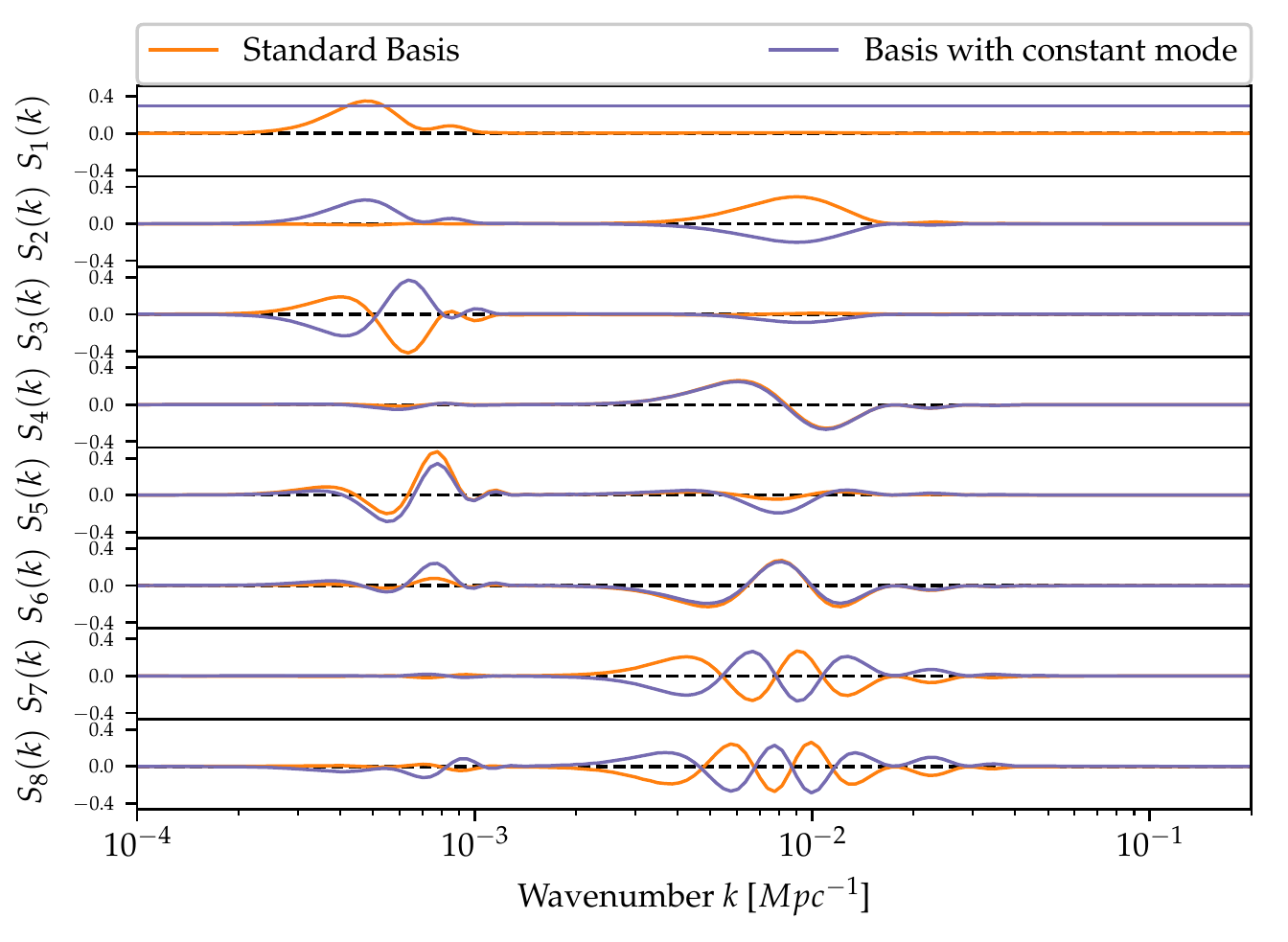}
	\caption{The first 8 PCA modes $\mathcal{S}_a(k)$ for the LiteBIRD experiment in the standard basis (solid orange curve) and in the constant $r$ mode basis (solid purple curve). Note that the first mode of the constant $r$ mode basis has been arbitrarily rescaled here to 0.3 to be shown in the same figure as the standard basis, but in reality is normalized to 1, as explained in the text. \label{fig:pca_modes_LB_with_without_ortho}}
\end{figure}

To start with, let us highlight briefly the consequences of including $r$ in the parametrization of $\mathcal{P}_T$ (and in the set of cosmological parameters against which we orthogonalize the PCA modes).
Clearly, we can equivalently impose that in our tensor power spectrum basis $\{\mathcal{S}_a\}$ we have $\mathcal{S}_1=(1,...,1)$, so that $m_1$ is effectively $r$ and the associated uncertainty is $\sigma_{r}$ (we refer to this choice as the \textit{basis with the constant mode} and contrast it with our \textit{standard basis} described in Sections \ref{sec:litebird} and \ref{sec:SOandS4}). The $\mathcal{S}_a$ functions for $a > 1$ are still PCA modes, but the orthogonality with $\mathcal{S}_1$ changes their shape. Most notably, $\mathcal{S}_1$ can be the only positive definite mode, forcing all the others to be oscillatory -- as can be seen in Figure \ref{fig:pca_modes_LB_with_without_ortho} for the LiteBIRD configuration\footnote{Incidentally, we note that the orthogonalization with respect to the other six cosmological parameters \{$A_{s}$, $n_{s}$, $\tau$, $\Omega_{b}h^{2}$, $\Omega_{D}h^{2}$, $H_{0}$ \} has no effect either on the shapes or the uncertainties of the modes. The reason is that most of the information about the modes come from the BB angular power spectrum, while the one about the six $\Lambda$CDM parameters come from TT, EE and TE.}. Note also that, in this constant mode basis, the tensor spectrum used to build the PCA basis has the same value of $r$ for which the basis is going to be used.

\begin{figure}
    \centering
    \includegraphics[scale=0.8]{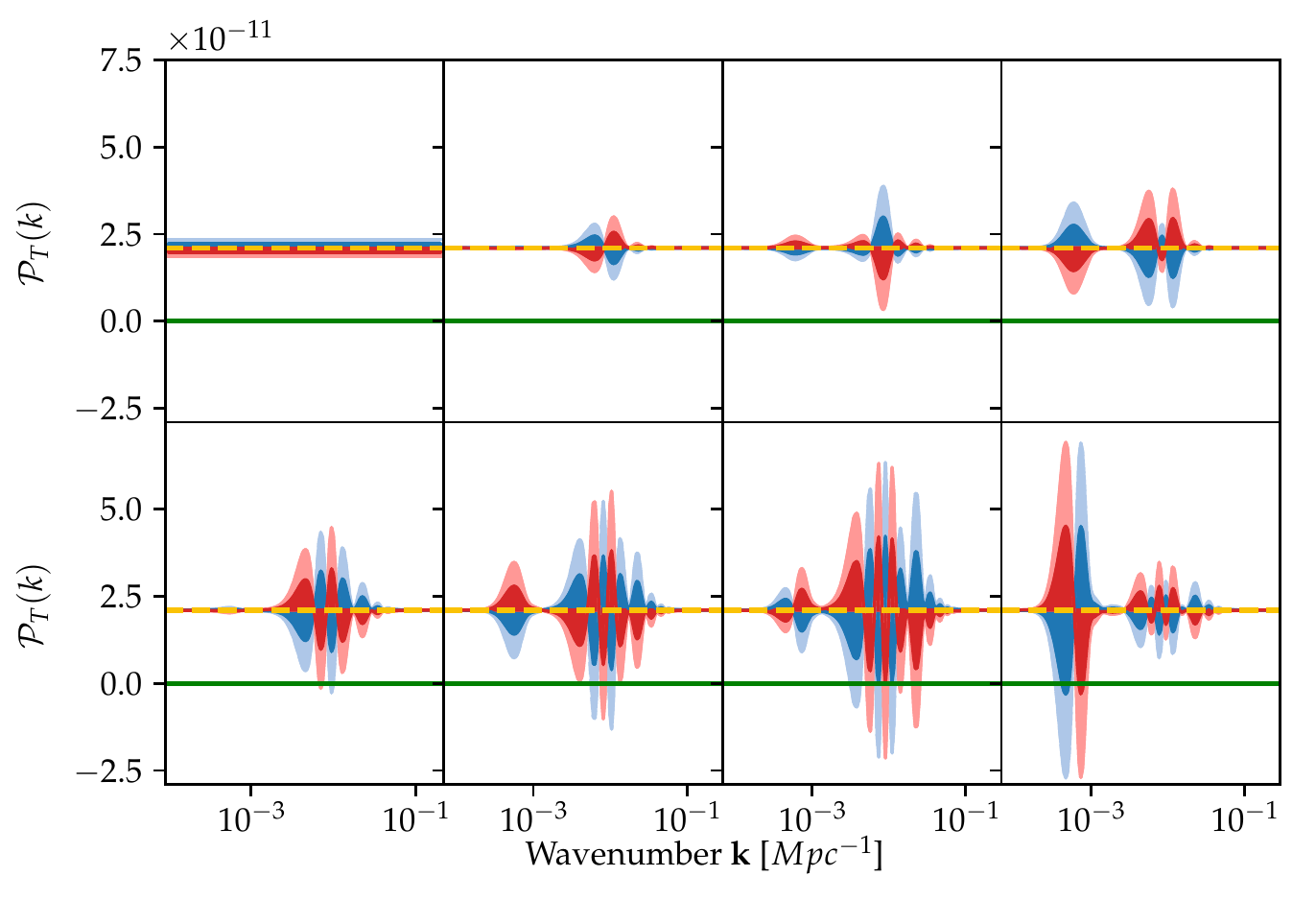}
\caption{Plots of $(r A_{s} \pm A_{s} \, \sigma_{a, Fisher} \,\mathcal{S}_a(k))$  for each mode $a$, for the first eight PCA modes in the constant mode basis with $r=0.01$ for LiteBIRD (dark blue for $+$ and dark red for $-$ shaded contours). The light blue (light red) shaded represents show the $+2\sigma$ ($-2\sigma$) contours and the yellow dashed line represents $A_{T}$. The top panel contains the first four modes in increasing order from left to right and the bottom panels the modes 5-8. The physicality prior is indicated with a solid green line.}\label{fig:sigma_plot}
\end{figure}

\begin{figure}
    \centering
    \includegraphics[scale=0.8]{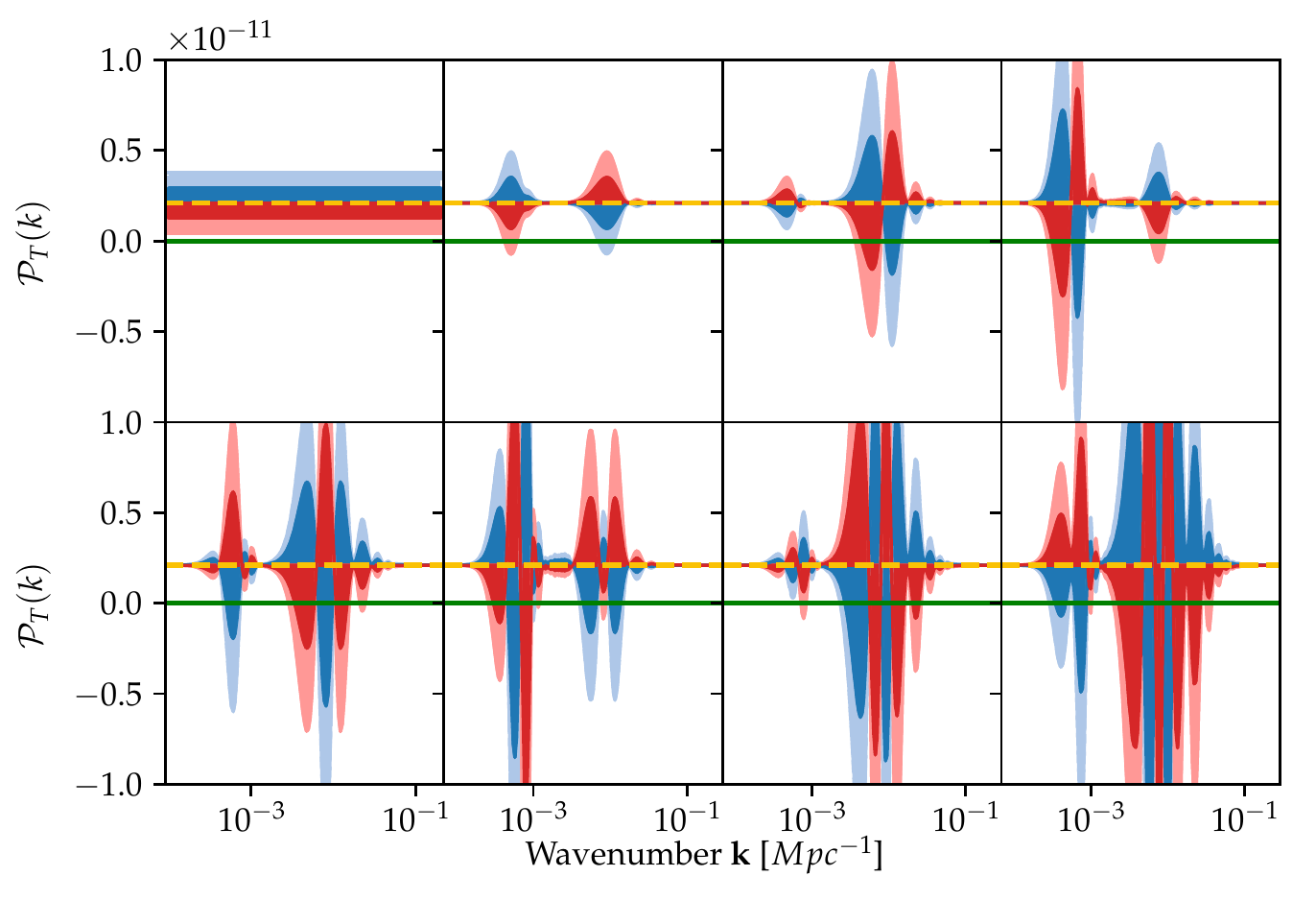}
\caption{Plots of $(r A_{s} \pm A_{s} \sigma_{a, Fisher} \mathcal{S}_a(k))$ for each mode $a$, for the first eight PCA modes in the constant mode basis with $r=0.001$ for LiteBIRD. 
}\label{fig:sigma_plot_0001}
\end{figure}

We now discuss a first visual argument showing that the Fisher estimates are very often inconsistent with the physicality prior. We run the Fisher analysis on the LiteBIRD configuration using the basis with the constant mode and two input angular power spectra, generated with $r = 0.01$ and $r = 0.001$. Assuming that the $\sigma_a$ we obtain are correct, we plot in Figure \ref{fig:sigma_plot} and \ref{fig:sigma_plot_0001} the $2\sigma$ range in which the modes are statistically expected to oscillate in 95\% of the cases. When $r=0.01$, for the modes up to $\mathcal{S}_{4}$ this range does not intersect the physicality prior. For the other modes, the intersection means that the Fisher estimate is unrealistic, as it predicts an oscillation larger than what is allowed by the physicality prior.
The case $r=0.001$, due to the lower $S/N$, is even more affected by the prior (Figure \ref{fig:sigma_plot_0001}): the $2\sigma$-contours intersect the negative region for all the modes beyond $r$.

\begin{figure}
	\centering
	\includegraphics[scale=0.15]{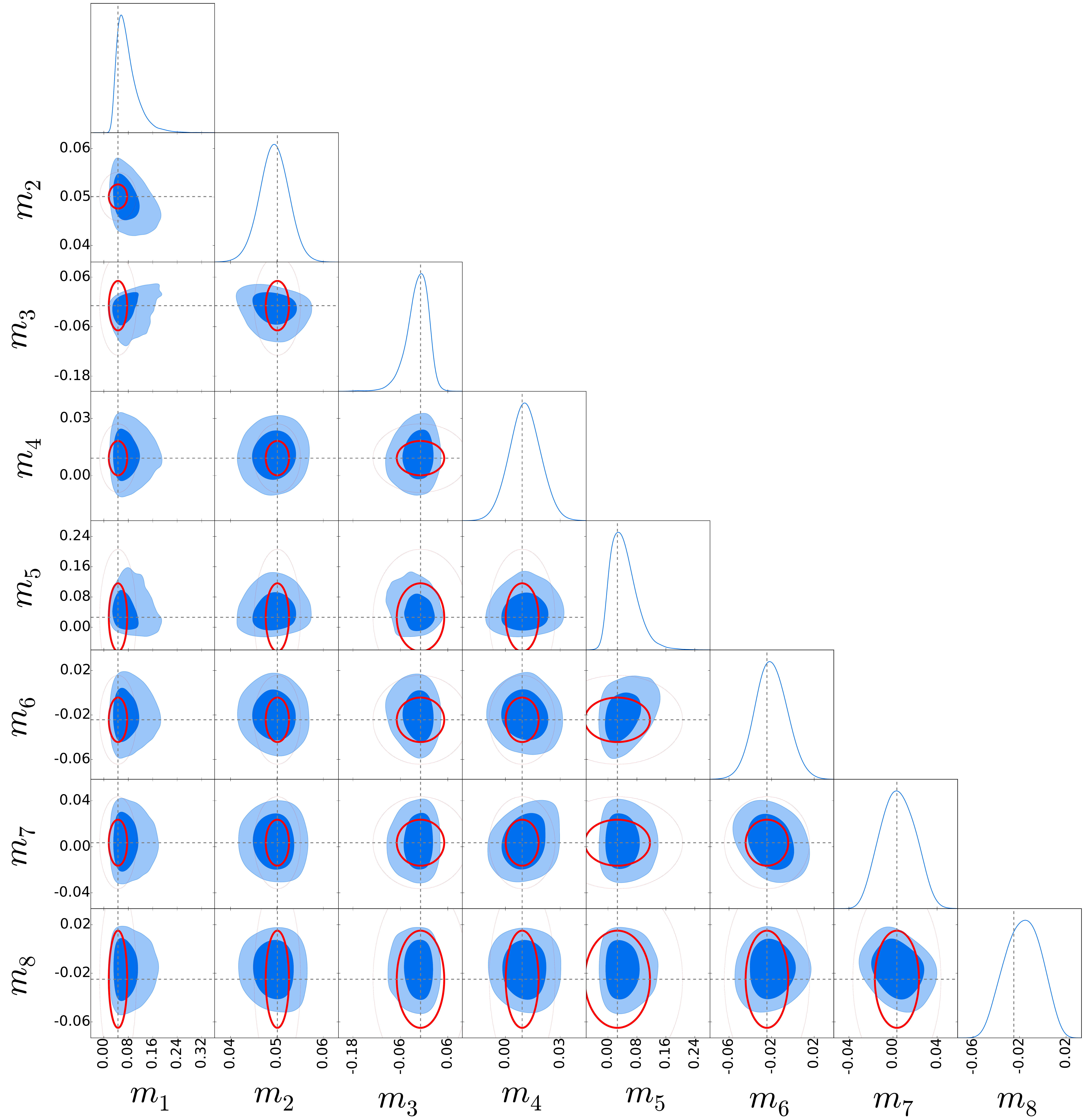}
	\caption{1D and 2D marginal distributions of the first eight PCA amplitudes for the standard PCA basis for LiteBIRD, using as input for the MCMC a model with $r=0.01$ (in blue). The contours represent 68\% and 95\% CL and the dashed grey line is showing the fiducial values of $m_a$. Also shown are the contours expected from the Fisher analysis (thick red ellipses for the $1\sigma$ errors and thin ones for $2\sigma$).
		\label{fig:standard_basis}}
\end{figure}

\begin{figure}
	\centering
	\includegraphics[scale=0.15]{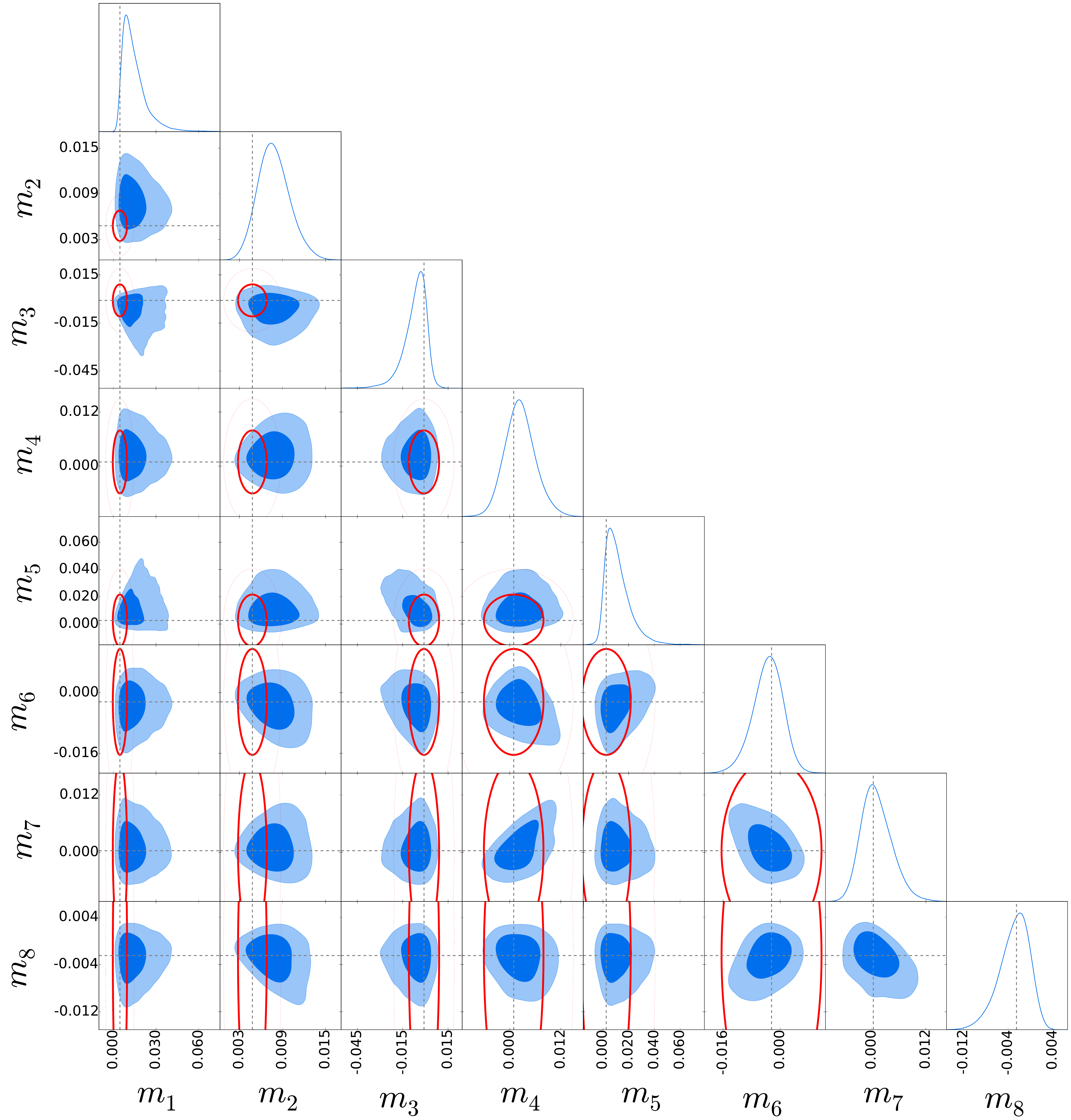}
	\caption{1D and 2D marginal distributions of the first eight PCA amplitudes for the standard PCA basis for LiteBIRD, using as input for the MCMC a model with $r=0.001$. 
		\label{fig:polygon}}
\end{figure}

\begin{figure}
	\centering
	\includegraphics[scale=0.15]{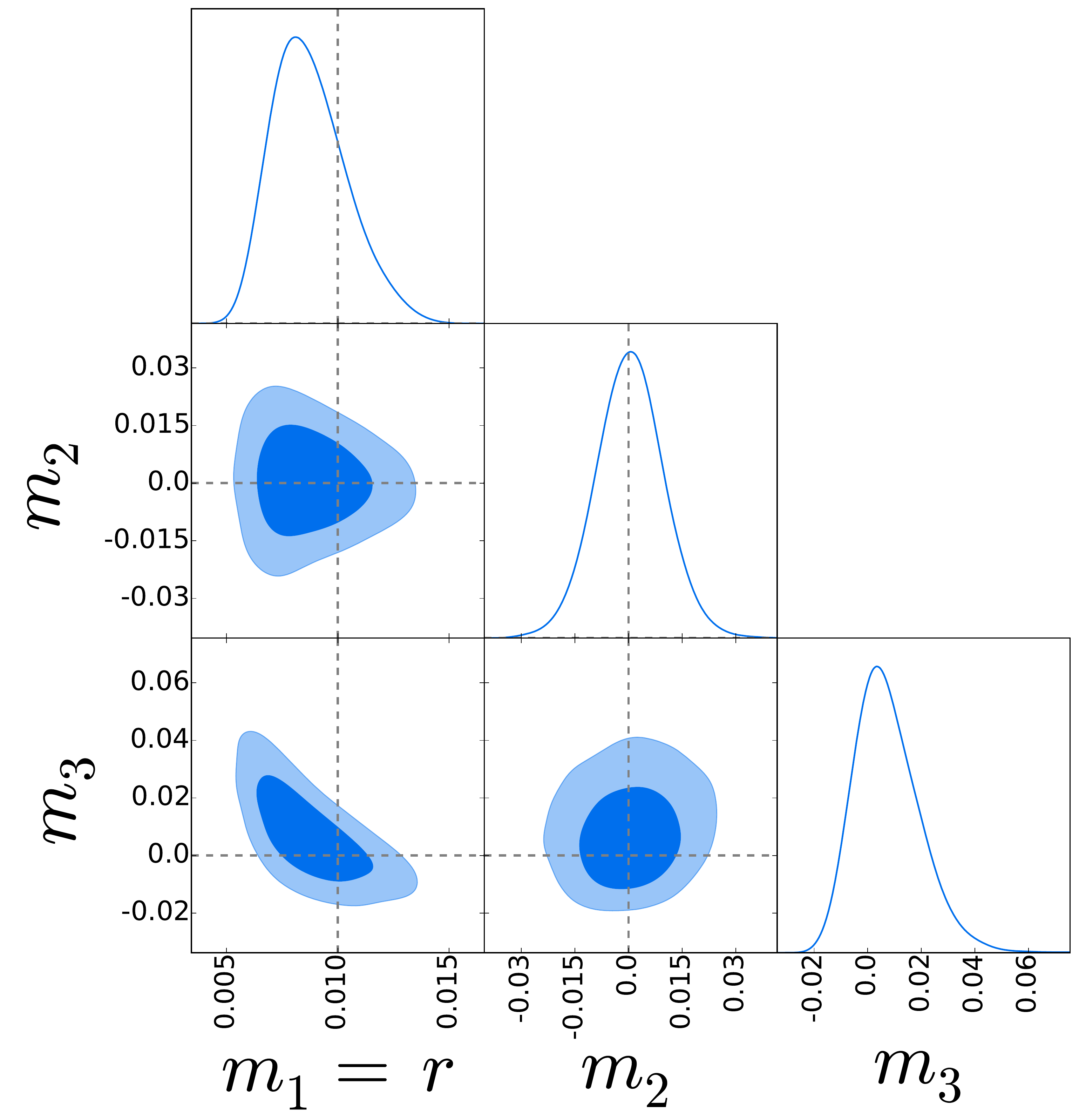}
	\caption{1D and 2D marginal distributions of the first three PCA amplitudes in the constant mode basis with $r=0.01$ for LiteBIRD. Note that in this basis the first mode has $m_1=r$. 
		\label{fig:amplitudes}}
\end{figure}

Second, we now run actual MCMC to estimate $\{m_a\}$ and the other cosmological parameters. We consider
input power spectra generated with $r = 0.01$ and $r = 0.001$, all the three experimental configurations and basis both with and without constant mode. The uncertainties derived from the MCMC estimation $\sigma_{MCMC}$ have to be larger or equal to the Fisher ones $\sigma_{Fisher}$. Without the physicality prior we checked that it is indeed the case, but as we turn on the prior the constraints that it imposes on the $m_a$ values often dominate or are comparable with those imposed by the observation, resulting in $\sigma_{Fisher} > \sigma_{MCMC}$ for most of the modes (Tables \ref{table:litebird_mcmc_vs_fisher}-\ref{table:cmbs4_mcmc_vs_fisher}). Only very few of the highest S/N modes satisfy $\sigma_{Fisher} < \sigma_{MCMC}$ and, interestingly, in the LiteBIRD case, for $r = 0.001$, their number is two when we use our standard basis and only one when we impose a constant mode. The effect of the physicality priors is visible even more clearly in the 1D and 2D marginal distribution of the MC samples (Figures \ref{fig:standard_basis} and \ref{fig:polygon}): the marginal distributions are strongly asymmetric, the contours have often polygonal shapes and are very different from the ones expected from the Fisher analysis (red ellipses). As it is evident from these plots, the non-Gaussianity of the contours is such that the maxima of the marginal distributions are significantly different from the best-fit values (which of course match the input values of the parameters). As a side comment, in the LiteBIRD case and input power spectrum with $r=0.01$, we compare our standard basis and the basis with the constant mode (only the first three modes, figure \ref{fig:amplitudes}). We note that the latter case shows a $m_1$-$m_3$ correlation\footnote{We can understand this 2D marginal as follows. Figure \ref{fig:sigma_plot} shows that $\mathcal{S}_3$ is strongly peaked in a very small $k$-range and, consequently, $m_a \mathcal{S}_a$ can hit the prior even for $m_3$ very close to zero, if they are negative. This prior effect occurs more easily when $m_1$ (i.e. $r$) is small, producing the $m_1$-$m_3$ correlation.}, while the former has the kind of shape we expect from uncorrelated parameters with asymmetric probability distributions. This behaviour of our basis is preferable, as the PCA should ideally yield uncorrelated parameters.
\begin{table}
		\centering
	\footnotesize{
		\begin{tabular}{|c|c|c|c|c|c|}
			\hline \hline
			Experiment 
			& PCA basis construction
			& Input MCMC
			& PCA mode
			& $\sigma_{Fisher}$
			& $\sigma_{MCMC}$
			\\
			\hline	\hline
	                  &   &	& 1st & 0.03 & 0.04 \\
		              &   &	& 2nd & 0.003 & 0.004 \\
		&  &	& 3rd & 0.06 & 0.03  \\
	 	         &   &      r=0.01	& 4th & 0.009 &  0.009 \\
	  	  &  &	& 5th & 0.09 & 0.04 \\
		   &          &    	& 6th & 0.02 &  0.014 \\
	          &      &		& 7th & 0.02 &  0.014 \\
	               & Standard &		& 8th & 0.04 &  0.014  \\
			\cline{3-6}
	              & basis     &  	& 1st & 0.005 & 0.009 \\
		          &     &  	& 2nd & 0.002 & 0.002 \\
		&  &	& 3rd & 0.01 & 0.008 \\
		&   &  r=0.001  	& 4th & 0.007 & 0.004 \\
	 &	 &    	& 5th & 0.019 & 0.01 \\
		&          &       	& 6th & 0.014 & 0.004 \\
		   &        &      	& 7th & 0.019 & 0.003\\
\textbf{LiteBIRD}	       &         & 		& 8th & 0.03 &  0.003\\
		 				\cline{2-6}
		 		 &      &    & 1st (const.) & 0.0006 & 0.0017  \\
			       &    &    & 2nd & 0.007 &  0.01 \\
			        &    &   & 3rd & 0.01 & 0.012 \\
		        & & r = 0.01 & 4th & 0.019  & 0.015 \\
	      &       & & 5th & 0.019 &  0.016 \\
	              &  &       & 6th & 0.03 & 0.017  \\
		           &     &   & 7th & 0.04  & 0.019 \\
			    & Constant mode    &      & 8th & 0.04 &  0.017 \\
				   	\cline{3-6}
				&  basis  &       & 1st (const.) & 0.0004 & 0.0005 \\
			     &    &      & 2nd & 0.003 & 0.002  \\
			     &    &      & 3rd & 0.007 & 0.002  \\
		         &   & r = 0.001 & 4th & 0.009  & 0.0017 \\
	    &   &  & 5th & 0.014 &  0.002 \\
	             &    &     & 6th & 0.019 &  0.0016 \\
		         &       &   & 7th & 0.019  &  0.002 \\
			    &     &      & 8th & 0.03 &  0.002 \\
				   \hline\hline

		\end{tabular}}
	\caption{Comparison between uncertainties on the first eight PCA modes for LiteBIRD for different PCA bases and different values of $r$. In the first two cases we take our PCA basis produced with $r=0$ (standard basis)  and project on it the models with $r=0.01$ (first case from the top) and $r=0.01$ (second case from the top). In the third and fourth case from the top instead, we construct the PCA basis with a constant $r=0.01$ and $r=0.001$ mode, respectively, so that $m_1 = r$. The third column from the left shows the value of $r$ chosen in the simulation used as input data for the MCMC.}
	\label{table:litebird_mcmc_vs_fisher}
\end{table}

\begin{table}
		\centering
	\footnotesize{
		\begin{tabular}{|c|c|c|c|c|c|}
			\hline \hline
			Experiment 
			& PCA basis construction
			& Input MCMC
			& PCA mode
			& $\sigma_{Fisher}$
			& $\sigma_{MCMC}$
			\\
			\hline	\hline
	& &		& 1st & 0.006 & 0.007 \\
	& &	$r=0.01$			& 2nd & 0.014 & 0.014 \\
		& &			& 3rd & 0.03 & 0.019 \\
		& Standard &			& 4th & 0.04 & 0.02 \\
			\cline{3-6}
		& basis &			     &  	1st & 0.004 & 0.004 \\
		& &	$r=0.001$		& 2nd & 0.01 &  0.006\\
		& &			& 3rd & 0.019 & 0.007 \\
\textbf{SO}		& &			& 4th & 0.03 &  0.008\\
					       	\cline{2-6}
			& &		       	& 1st (const.) & 0.0013 & 0.002  \\
			      &  & $r=0.01$		   & 2nd & 0.014 & 0.014  \\
			 & &		        & 3rd & 0.019 & 0.017 \\
			& Constant mode &		     & 4th &  0.04 & 0.017 \\
				   \cline{3-6}
			& basis&			    & 1st (const.) & 0.0009 & 0.0009 \\
			     &  & $r=0.001$		    & 2nd & 0.01 & 0.0019  \\
			 &  &		        & 3rd & 0.019 & 0.002  \\
			&  &		      & 4th &  0.03 & 0.0018 \\
				   \hline\hline

		\end{tabular}}
	\caption{Comparison between uncertainties on the first four PCA modes for SO for different PCA bases and different values of $r$. }
	\label{table:so_mcmc_vs_fisher}
\end{table}

\begin{table}
		\centering
	\footnotesize{
		\begin{tabular}{|c|c|c|c|c|c|}
			\hline \hline
			Experiment 
			& PCA basis construction
			& Input MCMC
			& PCA mode
			& $\sigma_{Fisher}$
			& $\sigma_{MCMC}$
			\\
			\hline	\hline
	& &		 &	1st & 0.005 & 0.008 \\
	& &		& 2nd & 0.014 & 0.014 \\
	& &	$r=0.01$	& 3rd & 0.02 & 0.02 \\
	& &		& 4th & 0.03 & 0.02 \\
	& &		& 5th & 0.04 & 0.021 \\
	& Standard &		& 6th & 0.05  & 0.022 \\
			\cline{3-6}
	& basis &	    & 1st & 0.002 & 0.003 \\
	& &		& 2nd & 0.006 & 0.004 \\
	& & $r=0.001$	& 3rd & 0.01 & 0.0057 \\
	& &		& 4th & 0.019 & 0.0059 \\
	& &		& 5th & 0.01 & 0.0061 \\
\textbf{CMB-S4}	& &		& 6th & 0.03  & 0.006 \\
				   \cline{2-6}
	& &			    & 1st (const.) & 0.0012 & 0.0019  \\
	& &			   	 & 2nd & 0.01 &  0.014 \\
	&   &	$r=0.01$	         & 3rd & 0.014 & 0.014  \\
	& &	   & 4th & 0.03  & 0.015 \\
	 & & & 5th & 0.03 & 0.015  \\
	   & Constant mode &           & 6th & 0.04 &  0.016 \\
             \cline{3-6}
         & basis &            & 1st (const.) & 0.0005 & 0.0005  \\
			& &         & 2nd & 0.006 & 0.0019  \\
			   & & $r=0.001$     & 3rd & 0.01 & 0.002  \\
		 & &  & 4th & 0.019  & 0.0017 \\
 & &	  & 5th & 0.019 &  0.0016 \\
	    & &             & 6th & 0.03 &  0.0015 \\
				   \hline\hline
		\end{tabular}}
	\caption{Comparison between uncertainties on the first six PCA modes for CMB-S4 for different PCA bases and different values of $r$. 
	}
	\label{table:cmbs4_mcmc_vs_fisher}
\end{table}

Concluding, while we can always use the PCA basis to model the primordial tensor power spectrum, the Fisher uncertainties -- that we obtain as a byproduct of the PCA -- are rarely accurate in absolute terms and should be used only for relative comparisons.

\section{Conclusions and prospects}\label{sec:conclusions}

We studied the principal component analysis applied to the tensor primordial power spectrum, with the goal of investigating its capability to detect, in a model-independent way, deviations from scale-invariance through the B-modes of the CMB polarization anisotropies. 
The PCA technique consists in diagonalizing the Fisher matrix and taking its eigenvectors to form a basis of uncorrelated modes, called PCA modes. The modes are ranked from the best to the worst measured ones according to the inverse of the square root of the eigenvalue associated to each mode, which represents the uncertainty on that mode. 

We derived constraints on the power spectrum parameters using the specifications from three future B-mode probes, namely LiteBIRD, SO and CMB-S4. We included the contributions of gravitational lensing by LSS, instrumental noise and -- most important -- the residuals of diffuse foregrounds (dust and synchrotron) following a foreground cleaning procedure. We found that residuals have a major impact on the analysis, and, in order to have realistic signal-to-noise ratio for the experiment considered, they must be included. Indeed, depending on the experiment and the value of $T/S$ ratio $r$ considered, adding foregrounds residuals can increase even by a factor $\sim 4$ the uncertainty on $r$ and on the PCA modes. Moreover, we found that the effect of foregrounds is relevant for both satellite and ground-based experiments. 

Then, through the shapes of these PCA modes and the uncertainty associated to each mode, we characterized the $k$-range for which each of the three experiments will be most sensitive to features in the power spectrum. In particular, LiteBIRD showed peaks of sensitivity, corresponding to the reionization and the recombination bump in the B-mode spectrum, consistent with \citep{hiramatsu_etal_2018}. Moreover, we found that the relative importance of the two peaks quickly shifts from the reionization peak to  the recombination one as we probe values of $r$ different from zero. For SO and CMB-S4, instead, the sensitivity is peaked on the recombination bump with CMB-S4 showing significantly smaller uncertainties on the first modes with respect to SO. 
 
Throughout our discussion, we have explained how the choices we have made in constructing our basis basis try to address the obstacles to the application of the PCA to the tensor power spectrum. Firstly, since it is still undetected,  we use a reference angular power spectrum with no tensor contribution for constructing the Fisher matrix. We devised, however, a procedure to choose the number of PCA modes retained that makes the basis well suited to analyze angular power spectra with substantial tensor contributions. Second, we remove $r$ from the parametrization of the tensor power spectrum, which is then expressed solely as a linear combination of PCA modes. We found indeed that including $r$ makes difficult for any other mode to have Fisher predictions that are robust with respect to the physicality prior  $P_{T}>0$. In any case, this prior remains the main limitation to the applicability of the PCA analysis to the tensor power spectrum, as highlighted in our comparison with the MCMC analysis.
In concluding, from a theoretical point of view, we would like to stress that our analysis can be applied to any scenario of Early Universe to quantify possible departures from the simplest inflationary scenarios, as we do in a toy example of red-tilted model. 
From the observational perspective, our analysis highlighted the complementarity of ultimate B-mode experiments over the next decade, probing large scales interested by the reionization power in the B-modes, uniquely accessed from space (LiteBIRD), and smaller ones from ground based probes (SO, CMB-S4), jointly looking at the degree scales, where the signal from cosmological gravitational waves is imprinted at recombination.

\acknowledgments
We thank Guglielmo Costa, Joanna Dunkley, Josquin Errard, Masashi Hazumi, Andrew Jaffe, Eiichiro Komatsu, Nicoletta Krachmalnicoff, Adrian Lee, Paolo Natoli, Radek Stompor and Nicola Vittorio for useful comments and discussions. We acknowledge use of the CMB4CAST, CAMB, COSMOMC and FGBUSTER codes. We acknowledge support from the COSMOS Network of the Italian Space Agency (cosmosnet.it), as well as from the INDARK Initiative of the National Institute for Nuclear Physics. This research was supported in part by the RADIOFOREGROUNDS project, funded by the European Commission's H2020 Research Infrastructures under the Grant Agreement 687312. We acknowledge the NERSC super-computing center in Berkeley and the Ulysses super-computer at SISSA for supporting numerical analyses in this work. This is not an SO Collaboration paper. 

\bibliographystyle{JHEP}
\bibliography{biblio}

\end{document}